\documentclass[conference]{IEEEtran}
\IEEEoverridecommandlockouts
\usepackage{cite}
\usepackage{amsmath,amssymb,amsfonts}
\usepackage{algorithmic}
\usepackage{graphicx}
\usepackage{textcomp}
\usepackage{xcolor}

\usepackage{url}
\usepackage{color}
\usepackage{bm}
\usepackage{amsfonts}
\usepackage{amsmath}
\usepackage{booktabs}
\usepackage{threeparttable}
\usepackage{verbatim}

\usepackage[ruled]{algorithm2e}
\usepackage{breakurl}
\usepackage{subcaption}

\def\BibTeX{{\rm B\kern-.05em{\sc i\kern-.025em b}\kern-.08em
    T\kern-.1667em\lower.7ex\hbox{E}\kern-.125emX}}
\begin{document}

\title{A Composite Predictive-Generative Approach to Monaural Universal Speech Enhancement}

\author{Jie Zhang, Haoyin Yan, and Xiaofei Li
\thanks{J. Zhang and H. Yan are with the National Engineering Research Center for Speech and Language Information Processing (NERC-SLIP), University of Science and Technology of China (USTC), 230027, Hefei, China. X. Li is with the School of Engineering, Westlake University, 310030, Hangzhou, China. (Email: jzhang6@ustc.edu.cn, hyyan@mail.ustc.edu.cn, lixiaofei@westlake.edu.cn) 
} 
\thanks{This work is supported by Anhui Province Major Science and Technology Research and Development Project (S2023Z20004). 
}
}

\maketitle

\begin{abstract}
It is promising to design a single model that can suppress various distortions and improve speech quality, i.e., universal speech enhancement (USE). Compared to supervised learning-based predictive methods, diffusion-based generative models have shown greater potential due to the generative capacities from degraded speech with severely damaged information. However, artifacts may be introduced in highly adverse conditions, and diffusion models often suffer from a heavy computational burden due to many steps for inference. In order to jointly leverage the superiority of prediction and generation and overcome the respective defects, in this work we propose a universal speech enhancement model called PGUSE by combining predictive and generative modeling. Our model consists of two branches: the predictive branch directly predicts clean samples from degraded signals, while the generative branch optimizes the denoising objective of diffusion models. We utilize the output fusion and truncated diffusion scheme to effectively integrate predictive and generative modeling, where the former directly combines results from both branches and the latter modifies the reverse diffusion process with initial estimates from the predictive branch. Extensive experiments on several datasets verify the superiority of the proposed model over state-of-the-art baselines, demonstrating the complementarity and benefits of combining predictive and generative modeling.
\end{abstract}


\section{Introduction}
\label{sec:intro}

\IEEEPARstart{R}{eal-world} speech recordings are inevitably contaminated by background noises, room reverberation, codec artifacts, and/or other distortion types, resulting in a degradation of perceptual quality and intelligibility. Speech enhancement (SE) aims to restore clean speech from contaminated recordings, which is an indispensable front-end for advanced speech-based applications, e.g., human-computer interaction, speech communication, remote conferencing~\cite{chen15o_interspeech,7807946,8683385}. Existing SE algorithms are usually task-oriented and separately customized for denoising~\cite{9413580}, dereverberation~\cite{10261213} or speech super-resolution (SR)~\cite{9335252}. Recent studies propose to consider multiple noise sources simultaneously by designing a universal SE (USE) framework~\cite{pascual19b_interspeech,9414721,serrà2022universalspeechenhancementscorebased,scheibler24_interspeech}, which aim to improve the speech quality under any degradation conditions with a single universal model. This would be more promising for applications than classic task-specific methods.

Since traditional statistics-based SE algorithms often suffer from the non-stationarity of acoustic scenes~\cite{zhang2023sdw}, data-driven learning-based methods become the mainstream in this field, which can learn the non-linearity between noisy and clean speech signals~\cite{8369155}. This mainstream can be roughly categorized into predictive  and generative approaches. Based on supervised learning, predictive methods usually treat SE as a regression task and learn the best mapping from degraded signals to the target signals under certain optimization criteria. 
Numerous studies focused on magnitude-level operations in the short-time Fourier transform (STFT) domain, since phase spectrum was somewhat unimportant for SE~\cite{1163920}. 
Subsequently, as it was shown that proper operations on phase can help to improve the perceptual quality and speech intelligibility~\cite{6891278},
complex-domain methods were then proposed by incorporating phase estimation, such as complex ratio masking~\cite{7364200} and complex spectral mapping~\cite{8168119}. 
To mitigate the compensation effect caused by penalizing real and imaginary parts, both complex and magnitude losses should be included for model training~\cite{9552504,10508391}.
Benefited from the development of deep neural network (DNN), time-domain approaches that directly operate on speech waveforms also show to be very promising in improving the speech quality~\cite{pascual17_interspeech,kim21h_interspeech}.

It is intuitive in the sense of USE that generative methods hold a greater potential because some distortions require the model to generate signals from scratch, e.g., clipping and bandwidth limitation~\cite{serrà2022universalspeechenhancementscorebased,pascual19b_interspeech,zhang21k_interspeech}.
These distortions involve irreversible information loss, offering challenges in predicting plausible reconstructions through deterministic mapping. Generative models aim to integrate the inherent distribution of data into latent space and generate samples from it, compensating for missing information via learned priors. Different from the predictive models providing unique prediction, generative methods have many possible candidates and allow stochasticity, which conforms to the various forms of signal reconstruction.
There are some typical examples. Variational auto-encoder (VAE)~\cite{vae} learns to represent data using an explicit probability distribution. Generative adversarial network (GAN)~\cite{10.5555/2969033.2969125} utilizes a discriminator to encourage the generator to generate realistic data. Normalizing flow~\cite{10.5555/3045118.3045281} employs a series of invertible transforms to obtain simple distributions transformed from complex data distributions. Diffusion models~\cite{10.5555/3045118.3045358,10.5555/3495724.3496298} simulate a probabilistic diffusion process, where data is gradually transformed into noise and the estimated signal is finally reconstructed as a reverse process. By constraining the generation step, one can deal with conditional generation tasks. The degraded 
speech can naturally be seen as the output of this conditioning, facilitating the applicability to USE~\cite{scheibler24_interspeech}.

Diffusion models have emerged as the new state-of-the-art (SOTA) family of deep generative models, especially for image generation~\cite{10.5555/3495724.3496298, 9878449, NEURIPS2021_49ad23d1}.
Recently, they have also been introduced to tackle SE and USE~\cite{9746901,welker22_interspeech,10149431,10180108,scheibler24_interspeech}. Diffusion models originally contain two parameterized discrete-time Markov chains, i.e., forward and reverse chains~\cite{10.5555/3495724.3496298}. The former gradually adds noise to the data until its distribution tends towards a tractable priori, which is usually the standard normal distribution. The latter learns to reverse this process and finally recovers the original distribution of data. By formalizing the diffusion process with stochastic differential equation (SDE), the discrete-time form can be converted into a continuous-time form~\cite{song2021scorebased}. Samples are generated by the score functions estimated at decreasing noise levels and using the score-based sampling approaches~\cite{JMLR:v6:hyvarinen05a}, which is called score-based diffusion model~\cite{song2021scorebased}. The resulting training and sampling operations are completely decoupled, allowing for flexible sampling strategies, and the continuous noise disturbance may lead to a smoother sample generation process~\cite{song2021scorebased}. We thus focus on score-based diffusion models in this work.

As rough attempts of integrating predictive and generative models, the stochastic refinement method~\cite{10095850}
utilizes a residual between the predictive output and the degraded speech for further diffusion. However, learning the residual is challenging due to its implicit structure and the low signal-to-noise ratio (SNR) at the start point of the reverse diffusion. The stochastic regeneration~\cite{10180108} adopts a predictive model to perform initial speech recovery, followed by a generative model to re-generate the final sample. Artifacts caused by the generative process can thus be reduced, as the initial recovery decreases the speech uncertainty.
However, unreliable predictive outputs will affect the generative results due to the cascade structure.
In~\cite{serrà2022universalspeechenhancementscorebased} a condition network is utilized to encode the degraded speech and guide the score estimation network, but there lacks an integration between predictive and generative results, which is crucial for improving speech reconstruction quality.
In addition, the heavy computational burden of diffusion models still exists since the reverse process requires numerous calls of the score estimation network, which hinders the applicability of diffusion-based SE models.
In contrast, predictive models perform the direct mapping or masking from degraded speech signals to the clean counterparts with single call, which is theoretically possible to accelerate the reverse process. The combination of predictive and generative approaches is thus promising in both improving the speech reconstruction quality and reducing the computational complexity.

In this work, we therefore propose a composite model called \textbf{PGUSE}, by jointly leveraging \textbf{P}redictive and \textbf{G}enerative modeling for \textbf{U}niversal \textbf{S}peech \textbf{E}nhancement.
The proposed model contains two parallel branches to perform predictive and generative modeling, respectively, where each branch comprises an encoder-decoder architecture. The sub-band downsampling-upsampling scheme helps capture band-aware features, and the dual-path recurrent attention module is designed as the bottleneck to model temporal and frequency dependencies efficiently. Interaction modules extract information from predictive branches to assist score estimation. In order to integrate predictive learning and generative learning effectively, we propose to utilize output fusion and a truncated diffusion scheme. Specifically, the former performs weighting between predictive and generative results in the spectral domain, and the latter adopts the predictive results to approximate latent variables in the reverse process, which can reduce the number of sampling steps. We utilize several datasets that cover multiple distortions to evaluate the effectiveness of the proposed method, including additive noise, reverberation, clipping, bandwidth limitation, etc. Extensive experimental results demonstrate a better SE capacity and robustness than other SOTA baselines. The reproducible code and audio examples are available online\footnote{https://hyyan2k.github.io/PGUSE}.

The remainder of this paper is organized as follows. Section II provides preliminaries of the score-based diffusion models. Section III describes the proposed approach. Section IV presents the experimental setup, followed by evaluation results in Section V. Finally, Section VI concludes this work.

\section{Score-based Diffusion Models}
In order to guide the reader, we present some preliminaries of the score-based diffusion models in this section.
By converting the discrete-time diffusion to the continuous-time form with SDE, they can be characterized by the forward process, the reverse process and the reverse sampling method.

\subsection{Forward Process}

The forward process gradually introduces noise to disturb the data distribution, including the mean and variance, which can be governed by the following SDE~\cite{song2021scorebased}:
\begin{align}
{\rm d} \bm{X}_t = \bm{f}(\bm{X}_t, t) {\rm d}t + g(t) {\rm d}\bm{w}, 0 \leq t \leq T, \label{sde}
\end{align}
where $\bm{X}_t$ denotes the latent variable at time $t$ and $\bm{w}$ a standard Wiener process. 
The diffusion process starts from $\bm{X}_0$ and ends at $\bm{X}_T$, where $\bm{X}_0$ is usually the clean waveform in the time domain or spectral coefficients in the STFT domain in the context of speech processing. 
Since the complex STFT spectrum can be represented as real and imaginary parts, this process is typically real-valued.
Functions $\bm{f}(\bm{X}_t, t)$ and $g(t)$ are referred to as the drift and diffusion coefficients, respectively. The former guides the mean drift of data, and the latter controls the amount of additive Gaussian white noise.

Some works tailor the SDE to SE tasks, which provide degraded signals as reconstruction clues~\cite{welker22_interspeech,10149431,lay23_interspeech,jukic24_interspeech,richter2024investigating}.
We summarize two forms of SDE here: Ornstein-Uhlenbeck with Variance Exploding (OUVE)~\cite{welker22_interspeech,10149431} and Brownian Bridge with Exponential Diffusion (BBED)~\cite{lay23_interspeech}.

\subsubsection{OUVE}
Following the notations in~\cite{lay23_interspeech}, the OUVE SDE is parameterized as
\begin{align}
\bm{f}(\bm{X}_t, t) &= \gamma (\bm{Y} - \bm{X}_t), \\
g(t) &= \sqrt{c} k^t,
\end{align}
where $\gamma, c, k \in \mathbb{R}_+$. The stiffness $\gamma$ controls the transition of the mean from $\bm{X}_0$ to $\bm{Y}$, i.e., from the clean speech to the degraded version. 
The parameters $c$ and $k$ schedule noise levels. Given the initial conditions, the closed-form solutions to the mean and variance of the state $\bm{X}_t$ are given by
\begin{align}
\bm{\mu}(\bm{X}_0, \bm{Y}, t) &=  e^{-\gamma t} \bm{X}_0 + (1 - e^{-\gamma t}) \bm{Y}, \\
\sigma^2(t) &= { c(k^{2t}-e^{-2\gamma t}) \over 2(\gamma+ {\rm log} (k)) }.
\end{align}
In case $t \rightarrow \infty$, the mean of $\bm{X}_t$ converges to $\bm{Y}$. However, the total diffusion amount $T$ is finite in practice (usually set to 1), resulting in a prior mismatch, which is quantified by the difference between $\bm{X}_T$ and $\bm{Y}$. This mismatch will cause an unavoidable bias in the subsequent reverse process.

\subsubsection{BBED}
\begin{figure}
    \centering
    \includegraphics[width=0.95\columnwidth]{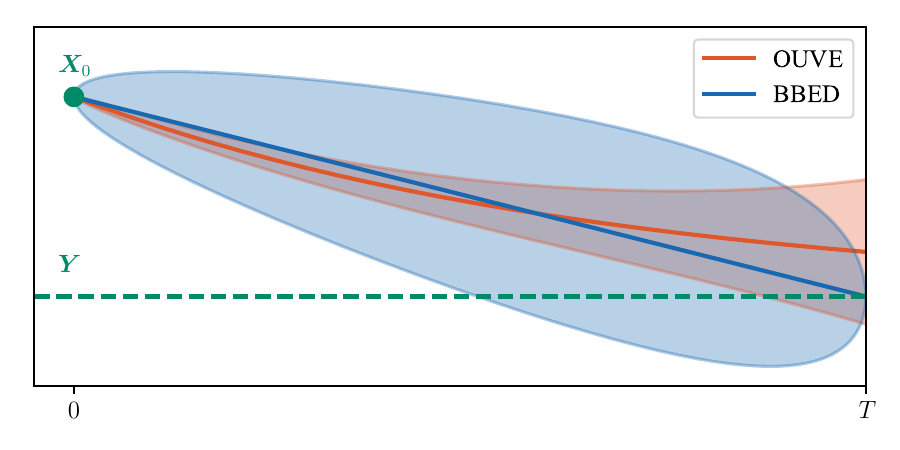}
    \vspace{-1em}
    \caption{Visualization of the forward process using OUVE and BBED SDE. Solid curves denote the mean value, and the variance is represented by the shaded area.}
    \label{fig:bbed}
    \vspace{-1em}
\end{figure}

The BBED improves the drift coefficient as
\begin{align}
\bm{f}(\bm{X}_t, t) = {\bm{Y} - \bm{X}_t \over 1 - t}, \label{drift}
\end{align}
with the diffusion coefficient being aligned with OUVE. This choice requires $T < 1$ due to the numerical stability. The mean and variance solution is determined by
\begin{align}
\bm{\mu}(\bm{X}_0&, \bm{Y}, t) =  (1-t) \bm{X}_0 + t \bm{Y},  \label{mean} \\
\sigma^2(t) &= (1-t)c \left[ (k^{2t}-1+t)+{\rm log}(k^{2k^2})(1-t)E\right], \\
E &= {\rm Ei}[2(t-1) {\rm log}(k) ] - {\rm Ei}[-2{\rm log}(k)],
\end{align}
where ${\rm Ei}[\cdot]$ denotes the exponential integral function. The forward processes of OUVE and BBED are depicted in Fig.~\ref{fig:bbed}. The drift in \eqref{drift} causes the linear mean evolution in~\eqref{mean}, and we see that the mean linearly approaches to $\bm{Y}$ when $t \rightarrow T$, resulting in a smaller prior mismatch if $T$ is close to 1 enough.

\subsection{Reverse Process and Sampling Method}

For any diffusion process in the form of  \eqref{sde}, it can be reversed by solving the following reverse SDE~\cite{ANDERSON1982313,song2021scorebased}:
\begin{align}
{\rm d} \bm{X}_t = [ -\bm{f}(\bm{X}_t, t) + g(t)^2 \nabla_{\bm{X}_t} {\rm log} p_t(\bm{X}_t) ] {\rm d}t + g(t) {\rm d} \bar{\bm{w}}, \label{reverse}
\end{align}
where $\bar{\bm{w}}$ is the standard Wiener process in the reverse-time flow. 
The solution trajectories of this reverse SDE exhibit the same marginal densities as those of the forward SDE, and the difference lies in evolving in the opposite time direction.
The gradient of the logarithmic probability density $\nabla_{\bm{X}_t} {\rm log} p_t(\bm{X}_t)$ is called score function~\cite{JMLR:v6:hyvarinen05a}. Given the initial condition $\bm{X}_0$ and ending state $\bm{Y}$, the latent variable $\bm{X}_t$ follows a Gaussian distribution as
\begin{align}
p_t(\bm{X}_t | \bm{X}_0, \bm{Y}) =  p_\mathcal{N}(\bm{X}_t;\bm{\mu}(\bm{X}_0, \bm{Y}, t), \sigma^2(t) \bm{I}),
\label{distri}
\end{align}
which is called perturbation kernel
with $p_\mathcal{N}$ being the probability density function of Gaussian distribution and $\bm{I}$ being a properly sized identity matrix. The score function is thus given by
\begin{align}
\nabla_{\bm{X}_t} {\rm log} p_t(\bm{X}_t | \bm{X}_0, \bm{Y}) = - { \bm{X}_t - \bm{\mu}(\bm{X}_0, \bm{Y}, t) \over \sigma^2(t) }.
\end{align}
Generating samples needs to solve \eqref{reverse}, but the score function is unavailable therein. For this, we can utilize a surrogate score model $s_\theta(\bm{X}_t, \bm{Y}, t)$ parameterized by a set of parameters $\theta$. The score model is optimized by minimizing the following denoising score matching objective~\cite{JMLR:v6:hyvarinen05a,6795935,song2021scorebased}:
\begin{align}
\mathcal{L}_{\rm score} = \mathbb{E}_{t, (\bm{X}_0, \bm{Y}), \bm{Z}} 
\left[ \left|\left| s_\theta(\bm{X}_t, \bm{Y}, t) + { \bm{Z} \over \sigma(t) } \right|\right|^2_2 \right], \label{dsm}
\end{align}
where $\bm{X}_t = \bm{\mu}(\bm{X}_0, \bm{Y}, t) + \sigma(t)\bm{Z} $ and $\bm{Z} \sim \mathcal{N}(\bm{0}, \bm{I})$ is randomly sampled during the training phase. 

During inference, the initial state of the reverse process $\bm{X}_T$ is sampled from $\mathcal{N}(\bm{Y}, \sigma^2(T) \bm{I})$. For OUVE, this sampling would cause a prior mismatch because the mean of $\bm{X}_t$ cannot reach $\bm{Y}$ with a finite $T$ during training. The mismatch can be reduced by increasing $T$ for fixed stiffness $\gamma$, which is equivalent to increasing $\gamma$ for fixed $T$~\cite{lay23_interspeech}. But increasing $\gamma$ will cause an unstable reverse process as discussed in~\cite{10149431}. For BBED, the sampling can match the training condition better, as shown in Fig. \ref{fig:bbed}. Using the score model $s_\theta$, the sample generation process can be performed by solving 
\begin{align}
{\rm d} \bm{X}_t = [ -\bm{f}(\bm{X}_t, t) + g^2(t) s_\theta(\bm{X}_t, \bm{Y}, t) ] {\rm d}t + g(t) {\rm d} \bar{\bm{w}}, \label{infer}
\end{align}
from $t=T \rightarrow 0$. 
The solution of \eqref{infer} depends on discrete time steps, which can be uniform, irregular or adaptive. In this work, we uniformly divide the interval $[0,T]$ into $N$ sub-intervals to discretize time steps. With the step size $\Delta t = T /N$, the reverse process is discretized as $\{ \bm{X}_T, \bm{X}_{T-\Delta t}, ...,  \bm{X}_0 \}$. There are many general-purpose numerical methods for solving SDEs, such as Euler-Maruyama and stochastic Runge-Kutta methods~\cite{numerical}. Special predictor-corrector samplers~\cite{song2021scorebased} combine numerical SDE solvers with score-based Markov Chain Monte Carlo approaches~\cite{PARISI1981378} to correct the marginal distribution of the estimated sample. Since there exists a corresponding deterministic process for any diffusion process, solving the probability flow ordinary differential equation (ODE) associated with \eqref{infer} also approximates the reverse process~\cite{song2021scorebased}. For simplicity, we will utilize the classic Euler-Maruyama method in this work.

\section{Proposed PGUSE Model}


\begin{figure*}
    \centering
    \includegraphics[width=1.0\textwidth]{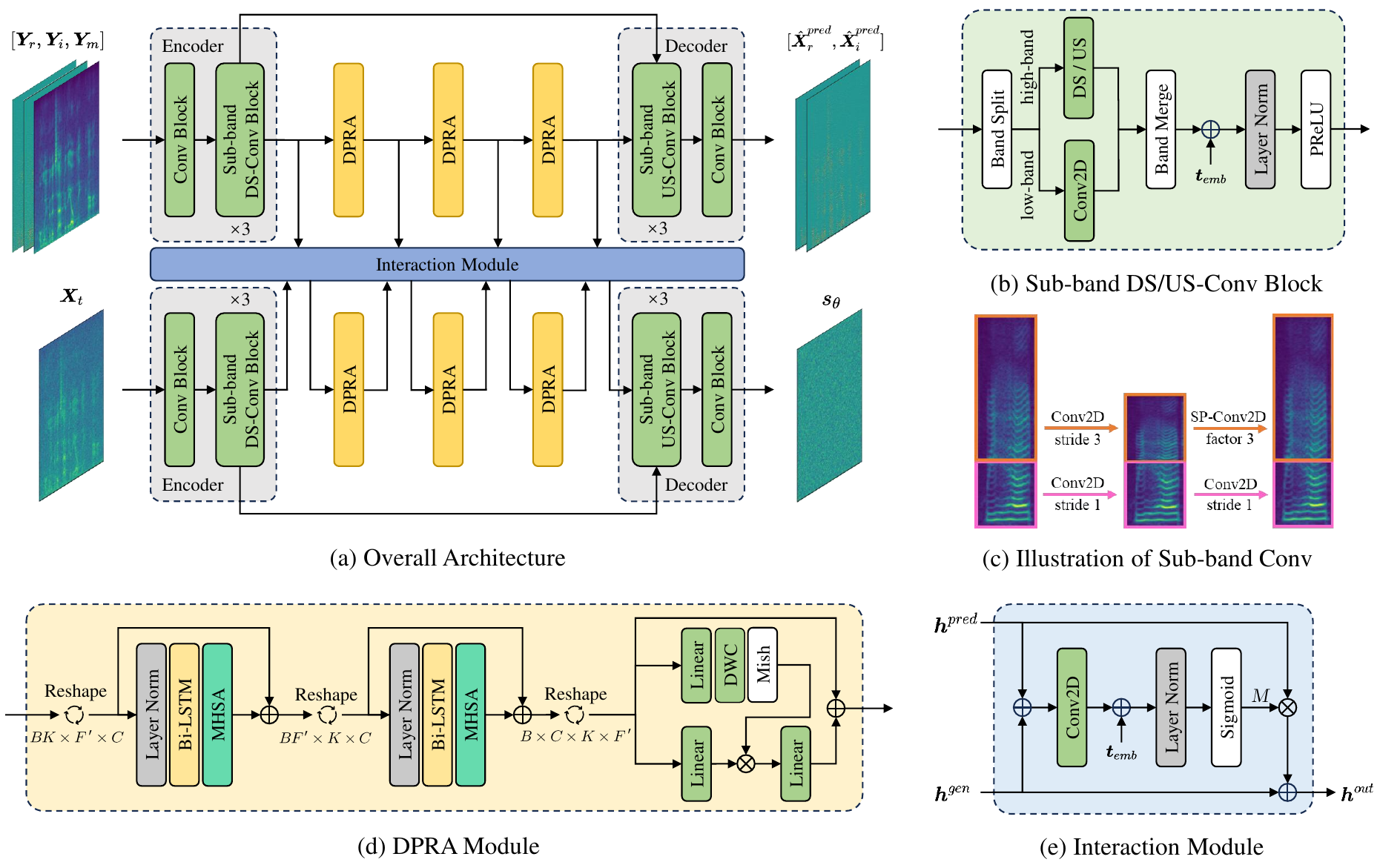}
    \caption{(a) The proposed PGUSE model, where the predictive branch (top) and the generative branch (bottom) are linked by the interaction module, (b) Sub-band DS/US-Conv block, (c) Sub-band Conv (from~\cite{yan2024lisennetlightweightsubbanddualpath}), (d) DPRA module with reshaping for frequency-temporal modeling and channel mixing, and (e) Interaction module by filtering data from the predictive branch to help estimate score functions.}
    \label{fig:architecture}
    \vspace{-1em}
\end{figure*}

\subsection{Data Representation}

Given the degraded speech waveform $\bm{y} \in \mathbb{R}^L$ caused by several distortions (e.g., additive noise, reverberation, clipping, bandwidth), the USE aims to restore the clean speech $\bm{x} \in \mathbb{R}^L$, where $L$ denotes the signal length. Instead of performing diffusion in the waveform domain~\cite{9746901,scheibler24_interspeech} or in the complex-valued STFT domain~\cite{10149431,10180108}, our generative method operates in the magnitude STFT domain. 
The additive Gaussian noise from diffusion process may lead to negative coefficients invalid for the magnitude spectrum~\cite{10149431}, but this issue can be addressed by clipping negative values to zero after reverse diffusion.
Compared to the waveform, the magnitude spectrum exhibits clear structures (e.g., formant), which are important for listening experience and are amenable for DNNs~\cite{Hermansky1990PerceptualLP,pmlr-v97-fu19b}. 
The complex spectrum map contains many unstructured textures in the image sense and challenges the score model's denoising process, which is typically operated by estimating the Gaussian noise at each diffusion state to approximate the score function as shown in \eqref{dsm}. This process becomes less effective when the spectral textures are not well-structured.
Therefore, the predictive branch of the proposed PGUSE model performs complex spectral mapping, in order to compensate for the lack of phase enhancement. 

Let $\tilde{\bm{Y}}\in\mathbb{C}^{K \times F}$ represent the complex spectrum (of the degraded signal) obtained by STFT, where $K$ and $F$ denote the number of time frames and frequency bins, respectively. 
Following~\cite{10149431,10508391}, we apply an amplitude transformation to each complex STFT coefficients $\tilde{\bm{Y}}(k,f)$ as
\begin{align}
\bm{Y}(k,f) = \beta_1 |\tilde{\bm{Y}}(k,f)|^{\beta_2} e^{i \angle \tilde{\bm{Y}}(k,f)} \label{transform},
\end{align}
where $\beta_1 \in \mathbb{R}_+$ is a scalar to roughly control the data range, $\beta_2 \in (0,1]$ is a compression exponent equalizing the importance of quieter sounds relative to loud ones, and $\angle \cdot$ denotes phase extractor. This transformation is not only meaningful in perceptual quality~\cite{1453591}, but also effective to enable networks to operate on a consistent data scale~\cite{9522648}.
In the sequel, all operations will be performed on transformed spectrum $\bm{Y}$ instead of the raw STFT matrix $\tilde{\bm{Y}}$.

\subsection{Network Architecture}

The overall architecture of our PGUSE model is depicted in Fig. \ref{fig:architecture}(a). The top part is the predictive branch, which takes the real part, imaginary part and magnitude component of the degraded speech as 3-channel real-valued input, say $\bm{Y}_r$, $\bm{Y}_i$ and $\bm{Y}_m \in \mathbb{R}^{K \times F}$, respectively. It performs complex spectral mapping to directly predict the real and imaginary parts of the clean speech, indicated as $\hat{\bm{X}}_r^{pred}$ and $\hat{\bm{X}}_i^{pred}$. The bottom half is the generative branch, estimating the score function from the current diffusion state $\bm{X}_t$. Note that the initial condition $\bm{X}_0$ and end state $\bm{Y}$ in the diffusion process are actually the clean magnitude spectrum $\bm{X}_m$ and the degraded counterpart $\bm{Y}_m$ in our implementation. The two branches almost share identical architectures modified from our previous work~\cite{yan2024lisennetlightweightsubbanddualpath}. The details of each module are discussed in the following.

\subsubsection{Encoder and Decoder}
The encoder consists of convolution (Conv) and sub-band downsampling convolution (DS-Conv) blocks, while the decoder includes several Conv and sub-band upsampling convolution (US-Conv) blocks. The Conv block is a cascade of a convolution layer, layer normalization (on the channel dimension) and parametric rectified linear unit (PReLU) activation. To reduce the computational complexity, the DS-Conv blocks progressively halve the frequency-axis size and maintain the time-axis size in the encoder, while the US-Conv blocks recover the frequency resolution in the decoder. 
Skip connections concatenate the outputs of DS-Conv blocks to the inputs of US-Conv blocks, facilitating the integration of low-level and high-level features~\cite{10.1007/978-3-319-24574-4_28}.
Generally, low-frequency bands contain more harmonic structures and play a more significant role in human hearing compared to high-frequency bands~\cite{Hermansky1990PerceptualLP}. We therefore utilize the sub-band Conv to extract band-aware features, see Fig. \ref{fig:architecture}(b) and \ref{fig:architecture}(c).
The feature map is initially split into low-band and high-band features. The former is processed by a 2D convolution (Conv2D) with a stride of 1 to maintain resolution, while the latter is downsampled using Conv2D with a stride of 3 (or upsampled via sub-pixel convolution (SP-Conv2D)~\cite{7780576} with a factor of 3) only along the frequency dimension, resulting in the same shape for both the low-band and downsampled high-band features. The outcomes are then concatenated as the full-band feature for subsequent processing.
To ensure that the model is time-dependent, Fourier-embeddings~\cite{NEURIPS2020_55053683,song2021scorebased} are employed to integrate time information of the diffusion process into the network. The scalar time index $t$ is mapped to vector $\bm{t}_{emb}$, which is added to the intermediate features before layer normalization. Since the predictive method is unrelated to the diffusion process, only the generative branch applies time embeddings, and therefore the two branches can be decoupled.

\subsubsection{Dual-Path Recurrent Attention (DPRA) Module}
It was recently demonstrated that dual-path~\cite{9054266} based models are effective for the SE task~\cite{10508391,le21b_interspeech,9746171}. We adapt the dual-path module to the diffusion-based SE model instead of directly transplanting from the image processing field~\cite{song2021scorebased,10149431,10180108}. Specifically, time sequences and frequency sequences are modeled sequentially to capture the time dependencies within each band and the frequency dependencies within each frame. The DPRA module is utilized as the bottleneck, which is extended from our previous work~\cite{yan2024lisennetlightweightsubbanddualpath} by incorporating the attention mechanism, e.g., see Fig. \ref{fig:architecture}(d). The hidden feature map $\bm{H} \in \mathbb{R}^{B \times C \times K \times F'}$ is initially reshaped to $BK \times F' \times C$ for frequency modeling and then reshaped to $BF' \times K \times C$ for temporal modeling, where $B$, $C$, and $F'$ denote the batch size, hidden dimension, and downsampled frequency-axis size, respectively. 
Each modeling process involves a cascade of layer normalization, bi-directional long short-term memory (Bi-LSTM), multi-head self-attention (MHSA)~\cite{NIPS2017_3f5ee243}, and residual connection. The MHSA attends to different positions in the sequence simultaneously, addressing the limitations of LSTM in retaining remote information. Subsequently, the convolutional gated linear unit (ConvGLU) modified from~\cite{10655567} serves as the channel mixer, which is composed of a linear layer, depthwise convolution (DWC)~\cite{8099678} and Mish~\cite{misra2019mish} non-linearity. The depthwise convolution aggregates the nearest information and the gate mechanism allows fine-grained channel attention.

\subsubsection{Interaction Module}
The relation between the generative and predictive branches is twofold: i) the predictive branch can provide clues of the degraded speech as the conditioning to guide the generation process; ii) the mapping process from the degraded speech to the clean signal can facilitate the denoising of the current diffusion state $\bm{X}_t$, since the theoretical mean of $\bm{X}_t$ depends on their interpolation as shown in \eqref{mean}. Based on this relation and inspired by~\cite{10064313}, we utilize an interaction module to transfer valuable supplementary information from the predictive branch to the generative branch. The interaction module is shown in Fig. \ref{fig:architecture}(e). The hidden feature $\bm{h}^{pred}$ from the predictive branch and $\bm{h}^{gen}$ from the generative branch are combined and fed into a cascade of Conv2D, layer normalization and Sigmoid activation to produce the mask $\bm{M}$, which is used to filter $\bm{h}^{pred}$. The time embedding $\bm{t}_{emb}$ is also employed here to introduce the time information. As a result, the output hidden feature $\bm{h}^{out}$ is given by
\begin{align}
\bm{h}^{out} = \bm{h}^{gen} + \bm{M} \otimes \bm{h}^{pred},
\end{align}
where $\otimes$ denotes the element-wise multiplication.

\subsection{Output Fusion}

The predictive methods often exhibit an over-suppression problem, i.e., speech components are excessively diminished during the denoising process~\cite{wang2020voicefilterlitestreamingtargetedvoice}. 
On the other hand, the generative approaches may introduce artifacts under highly adverse conditions, due to their inherent uncertainty regarding the presence or characteristics of speech~\cite{10180108}. In order to leverage the respective superiority and compensate the weakness, in this work we propose to perform output fusion given the predictive and generative results as 
\begin{align}
\hat{\bm{X}}_m &= \alpha \hat{\bm{X}}_m^{pred} + (1-\alpha) \hat{\bm{X}}_m^{gen}, \label{fusion}\\
\hat{\bm{X}}_m^{pred} &= \sqrt{ (\hat{\bm{X}}_r^{pred})^2 + (\hat{\bm{X}}_i^{pred})^2 }, \label{mag}
\end{align}
where $\hat{\bm{X}}_m^{pred}$ and $\hat{\bm{X}}_m^{gen}$ indicate the magnitude spectrums estimated by the predictive and generative branches, and $\alpha \in [0, 1]$ is the weighting factor. We perform magnitude-domain weighting, and the phase spectrum $\hat{\bm{X}}_p$ is simply taken from the predictive branch as 
\begin{align}
\hat{\bm{X}}_p = {\rm Arctan2}(\hat{\bm{X}}_i^{pred}, \hat{\bm{X}}_r^{pred}),  \label{pha}
\end{align}
where ${\rm Arctan2}$ is the two-argument arc-tangent function. The estimated magnitude and phase are finally coupled to restore the enhanced waveform by inverse STFT (iSTFT).

\subsection{Truncated Diffusion}

\begin{figure}
    \centering
    \includegraphics[width=0.95\columnwidth]{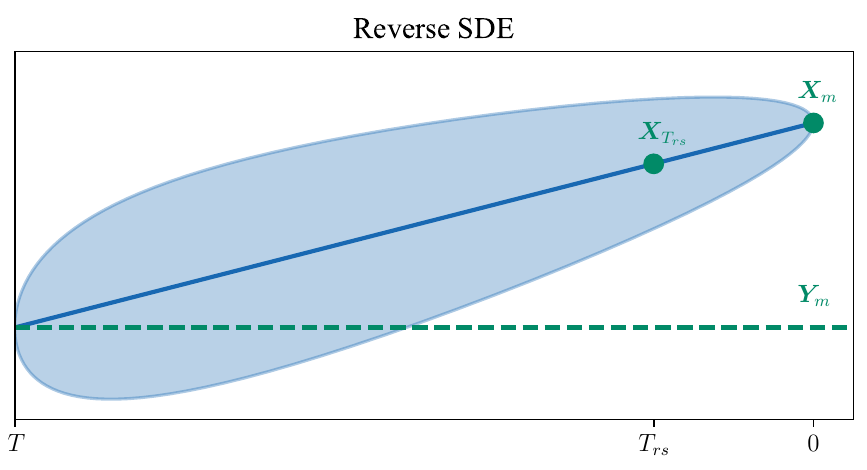}
    \vspace{-0.7em}
    \caption{Visualization of the reverse process of BBED SDE, where the truncated diffusion process starting from $T_{rs}$ with initial state $\bm{X}_{T_{rs}}$.}
    \label{fig:truncated}
    \vspace{-1em}
\end{figure}

Recently, some efforts were made to improve the sampling speed of the diffusion models by truncating the forward and reverse processes, called truncated diffusion~\cite{lyu2022acceleratingdiffusionmodelsearly,osti_10467093}. The key idea is to diffuse samples from pre-generated results instead of reversing the diffusion process from the Gaussian white noise, which can reduce the number of sampling steps. In this work, we make modifications and adapt the classic truncated diffusion scheme from discrete formulation to score-based diffusion process. We utilize the results from the predictive branch to accelerate sampling, leading to the exclusion of another model for pre-generated results. Specifically, we start the reverse process from $T_{rs} \in (0, T]$ instead of $T$. 
We replace the clean magnitude spectrum $\bm{X}_m$ with the predictive estimate $\hat{\bm{X}}^{pred}_m$ to approximate the theoretical mean of the initial state $\bm{X}_{T_{rs}}$, which is sampled from $\mathcal{N}(\bm{\mu}(\hat{\bm{X}}^{pred}_m, \bm{Y}_m, T_{rs}), \sigma^2(T_{rs}) \bm{I})$ and can be approximately given by
\begin{align}
\bm{X}_{T_{rs}} \approx \bm{\mu}(\hat{\bm{X}}^{pred}_m, \bm{Y}_m, T_{rs}) + \sigma(T_{rs}) \bm{Z}.
\label{trunc}
\end{align}
The visualization of truncated diffusion for BBED are shown in Fig. \ref{fig:truncated}. For a fixed step width $\Delta t$, a decreased diffusion time means fewer sampling steps, which thus saves the computational burden of diffusion models.

\subsection{Training Criteria}

For the predictive branch, we simultaneously consider the mean square errors (MSEs) on the magnitude spectrum and the complex spectrum as
\begin{align}
\mathcal{L}_{\rm mag} &= \mathbb{E}\left[ \left|\left| \hat{\bm{X}}_m^{pred} - \bm{X}_m \right|\right|_2^2 \right], \\
\mathcal{L}_{\rm comp} &= \mathbb{E} \left[ \left|\left| \hat{\bm{X}}_r^{pred} - \bm{X}_r \right|\right|_2^2 + \left|\left| \hat{\bm{X}}_i^{pred} - \bm{X}_i \right|\right|_2^2 \right].
\end{align}
The composite losses can help mitigate the compensation effect caused by only penalizing real and imaginary parts~\cite{9552504}. For the generative branch, we employ the denoising score matching objective given in \eqref{dsm}. Therefore, the overall loss function for the training of the proposed PGUSE is given by
\begin{align}
\mathcal{L} = \lambda \mathcal{L}_{\rm mag} + (1 - \lambda) \mathcal{L}_{\rm comp} + \mathcal{L}_{\rm score}, \label{loss}
\end{align}
where $\lambda$ balances the losses on the magnitude and complex spectrum, which is set to 0.5 in this work.


The training steps of our proposed PGUSE model are summarized in Algorithm~\ref{al1}, where the predictive and generative branches are represented by $P_\theta$ and $G_\theta$. Note that $\bm{h}^{pred}$ is not detached during training, such that the gradients from the generative branch can back-propagate to the predictive branch through the interaction module.
Algorithm~\ref{al2} outlines the inference process that adopts the classic Euler-Maruyama method to obtain the numerical solution of the reverse SDE. Since the predictive and generative branches can be decoupled, the repetitious calls of score estimating network are only involved in the generative branch.
The estimated real and imaginary components by Algorithm~\ref{al2} have to undergo the inverse transform of (\ref{transform}) and iSTFT to recover the time-domain enhanced waveform. 

\begin{algorithm}[t]
\SetAlgoLined
\KwIn{$\bm{X}_r$, $\bm{X}_i$, $\bm{X}_m$, $\bm{Y}_m$, $\bm{Y}_r$, $\bm{Y}_i$ }
\KwOut{$\mathcal{L}$}
\BlankLine
$[\hat{\bm{X}}_r^{pred}, \hat{\bm{X}}_i^{pred}]$, $\bm{h}^{pred} \leftarrow P_\theta([\bm{Y}_r, \bm{Y}_i, \bm{Y}_m])$ \;
Sample $t \sim \mathcal{U}(0, T)$ \;
Sample $\bm{Z} \sim \mathcal{N}(0, \bm{I})$ \;
$\bm{X}_t \leftarrow \bm{\mu}(\bm{X}_m, \bm{Y}_m, t) + \sigma(t) \bm{Z}$ \tcp*{\eqref{distri}}
$s_\theta \leftarrow G_\theta(\bm{X}_t, \bm{h}^{pred})$ \;
Calculate loss $\mathcal{L}$ using \eqref{loss}.
\caption{Training of PGUSE}
\label{al1}
\end{algorithm}

\begin{algorithm}[t]
\SetAlgoLined
\KwIn{$\bm{Y}_r$, $\bm{Y}_i$, $\bm{Y}_m$}
\KwOut{$\hat{\bm{X}}_r$, $\hat{\bm{X}}_i$}
\BlankLine
$[\hat{\bm{X}}_r^{pred}, \hat{\bm{X}}_i^{pred}]$, $\bm{h}^{pred} \leftarrow P_\theta([\bm{Y}_r, \bm{Y}_i, \bm{Y}_m])$ \;
$\hat{\bm{X}}_m^{pred} = \sqrt{ (\hat{\bm{X}}_r^{pred})^2 + (\hat{\bm{X}}_i^{pred})^2 }$ \tcp*{\eqref{mag}}
$\bm{X}_{T_{rs}} \leftarrow \bm{\mu}(\hat{\bm{X}}^{pred}_m, \bm{Y}_m, T_{rs}) + \sigma(T_{rs}) \bm{Z}$ \tcp*{\eqref{trunc}}

\For{$t=T_{rs}$, $T_{rs}-\Delta t$, $T_{rs}-2\Delta t$, ..., $\Delta t$}{
$s_\theta \leftarrow G_\theta(\bm{X}_t, \bm{h}^{pred})$ \;
Sample $\bm{Z} \sim \mathcal{N}(0, \bm{I})$ \;
$\bm{X}_{\rm mean} \leftarrow \bm{X}_t + (-\bm{f}(\bm{X}_t, t)+g(t)^2s_\theta) \Delta t$  \;
$\bm{X}_{t-\Delta t} \leftarrow \bm{X}_{\rm mean} + g(t) \sqrt{\Delta t} \bm{Z}$  \tcp*{\eqref{infer}}
}
$\hat{\bm{X}}_m^{gen} \leftarrow {\rm Clip}(\bm{X}_{\rm mean}, 0, +\infty)$ \;
$\hat{\bm{X}}_m \leftarrow \alpha \hat{\bm{X}}_m^{pred} + (1 - \alpha) \hat{\bm{X}}_m^{gen}$  \tcp*{\eqref{fusion}}
$\hat{\bm{X}}_p \leftarrow {\rm Arctan2}(\hat{\bm{X}}_i^{pred}, \hat{\bm{X}}_r^{pred})$ \tcp*{\eqref{pha}}
$\hat{\bm{X}}_r, \hat{\bm{X}}_i \leftarrow \hat{\bm{X}}_m{\rm cos}(\hat{\bm{X}}_p), \hat{\bm{X}}_m{\rm sin}(\hat{\bm{X}}_p)$ \;
\caption{Inference of PGUSE}
\label{al2}
\end{algorithm}

\section{Experimental Setup}
In this section, we will present datasets, evaluation metrics, implementation details, impacts of hyper-parameters as well as several SOTA comparison methods used in experiments.
\subsection{Datasets}

We utilize several datasets in experiments to evaluate the efficacy of our approach, with all audio samples sampled at 16 kHz. Each dataset is described below in detail.

\subsubsection{WSJ0-UNI}
We create the WSJ0-UNI dataset to evaluate our model on the USE task, incorporating multiple types of distortions. We employ the distortion pipeline\footnote{https://github.com/microsoft/SIG-Challenge} adapted from Speech Signal Improvement Challenge~\cite{10626457}, see Table \ref{tab:distortion}. The distortion families include recorded noise, reverberation, microphone frequency response, analog-to-digital converter (ADC) effects, automatic gain control (AGC) and transmission impacts. Specific distortions encompass additive noise, room impulse response (RIR) convolution, band filtering, bit depth adjustments, clipping, gain alterations, resampling, and data compression in global system for mobile communications (GSM). The clean speech utterances are sourced from the Wall Street Journal (WSJ0) dataset~\cite{wsj0} (distinct subsets ``si\_tr\_s", ``si\_dt\_05" and ``si\_et\_05" are used for training, validation and testing, respectively), while noise clips are randomly selected from WHAM! dataset~\cite{wichern19_interspeech}. 

\subsubsection{VBDMD}
We adopt the publicly available VoiceBank-DEMAND (VBDMD) dataset~\cite{valentinibotinhao16_ssw} for the denoising task, which is the often-used benchmark for monaural SE. The clean samples sourced from the VoiceBank corpus~\cite{6709856} contain 11,572 samples from 28 speakers for training and 872 clips from 2 speakers for testing. Clean signals are mixed with noises from the DEMAND dataset~\cite{10.1121/1.4806631} at SNRs of \{0, 5, 10, 15\} dB for training and \{2.5, 7.5, 12.5, 17.5\} dB for testing. All clips are resampled from 48 kHz to 16 kHz in experiments.

\subsubsection{VBDMD-REVERB}
Using clean utterances of the VBDMD test set, we apply the stereo reverb algorithm~\cite{1166351} similarly to WSJ0-UNI to generate the VBDMD-REVERB dataset, resulting in an average reverberation time (T60) of 0.4 seconds. This is used to evaluate the dereverberation capacity of USE models on unseen data.

\subsubsection{VBDMD-SR}
To evaluate the speech super-resolution (SR) (or bandwidth extension) ability of the model, we apply the 12-order Butterworth low-pass filter with a cutoff at 4 kHz to the clean samples of the VBDMD test set, resulting in the VBDMD-SR dataset. The speech SR requires models to extend frequency bands based on low-frequency information, which is a challenging generative task in the audio community.

\subsubsection{TIMIT-UNI}
We generate the TIMIT-UNI dataset using the same distortion pipeline as WSJ0-UNI, with clean speech utterances originated from the TIMIT corpus~\cite{timit}. Since the transcripts of TIMIT are available, we can then further evaluate the impact of SE models on the downstream automatic speech recognition (ASR) performance.

\begin{table}[]
    \small
    \centering
    \caption{Distortion categories and corresponding probabilities, grouped by family.}

    \resizebox{0.95\columnwidth}{!}{
        \begin{tabular}{lcc}
        \toprule
        \textbf{Family} & \textbf{Type} & \textbf{Probability} \\
        \midrule
        \midrule
        Noise & Additive noise & 0.3 \\
        \midrule
        Reverberation & RIR convolution & 0.25 \\
        \midrule
        Microphone & Low shelf filter & 0.5 \\
                   & High shelf filter &    \\
                   & Peak filter &          \\
        \midrule
        ADC & Low pass filter & 0.7 \\
            & High pass filter & 0.7 \\
            & Bit depth & 0.1 \\
        \midrule
        AGC & Clipping & 0.4 \\
            & Gain &                   \\
        \midrule
        Transmission & Clipping & 0.25 \\
                     & Gain & 0.25 \\
                     & Resample & 0.4 \\
                     & GSM compression & 0.25 \\
        \bottomrule
        \end{tabular}
    }
    \vspace{-1em}
    \label{tab:distortion}
\end{table}

\begin{figure*}[htbp]
    \centering
    \begin{subfigure}{0.3\textwidth}
        \centering
        \includegraphics[width=\linewidth]{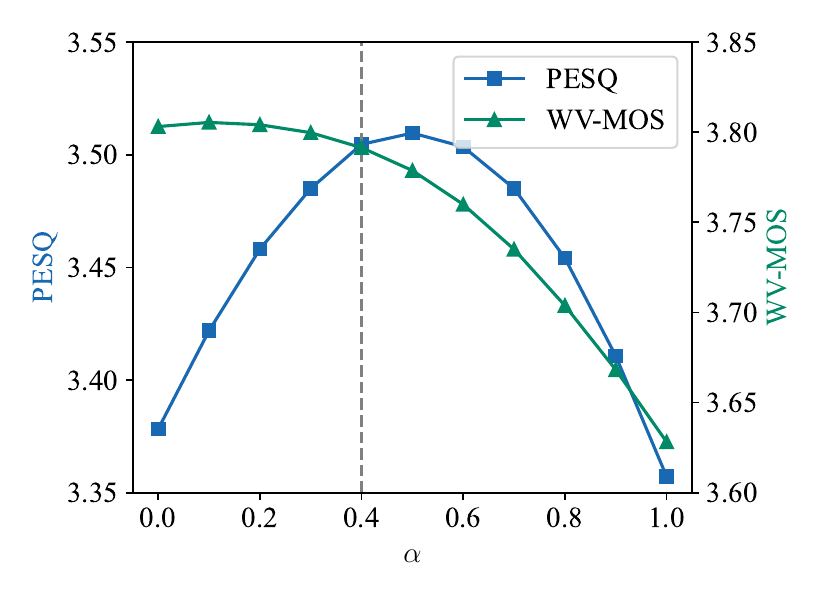}
        \caption{Varying $\alpha$ with $N=25$, $T_{rs}=T$}
        \label{fig:alpha}
    \end{subfigure}
    \centering
    \begin{subfigure}{0.3\textwidth}
        \centering
        \includegraphics[width=\linewidth]{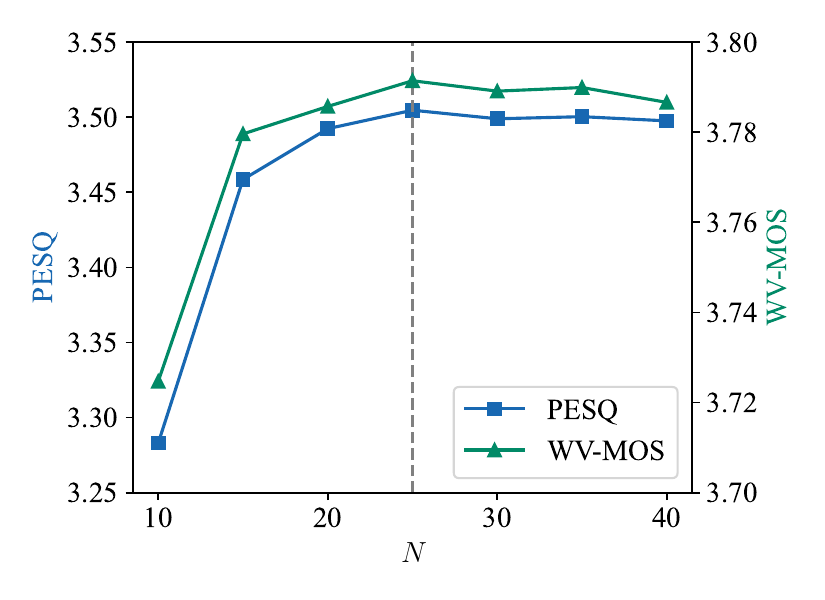}
        \caption{Varying $N$ with $\alpha=0.4$, $T_{rs}=T$}
        \label{fig:N}
    \end{subfigure}
    \centering
    \begin{subfigure}{0.3\textwidth}
        \centering
        \includegraphics[width=\linewidth]{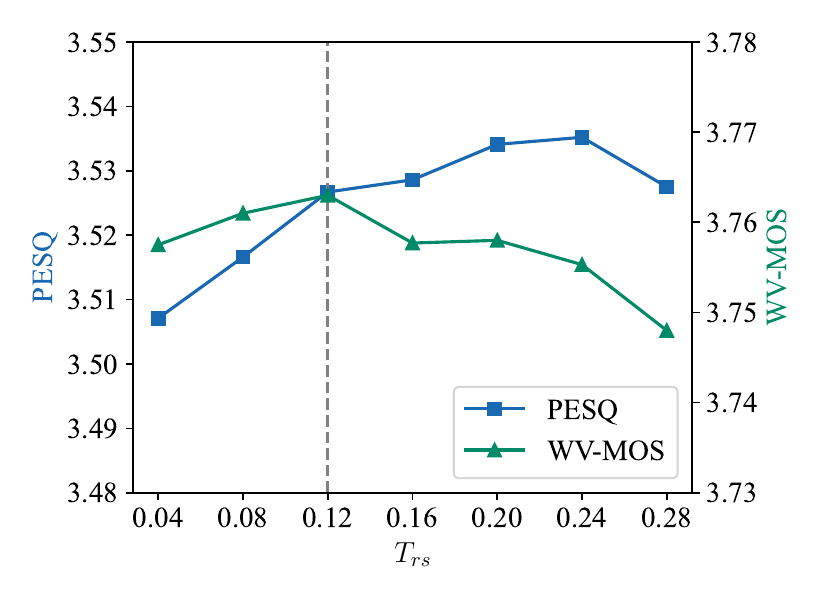}
        \caption{Varying $T_{rs}$ with $\alpha=0.4$, $\Delta t=0.04$}
        \label{fig:Trs}
    \end{subfigure}
    \caption{The performance analysis under different conditions of hyper-parameters.}
    \vspace{-1em}
    \label{fig:alpha_N_Trs}
\end{figure*}

\subsection{Evaluation Metrics}

In this work, we utilize several performance metrics for the instrumental evaluation of the proposed method.

\subsubsection{PESQ}
The perceptual evaluation of speech quality (PESQ)~\cite{941023} is widely-used for the objective speech quality evaluation. We employ the wideband PESQ scoring from 1 (poor) to 4.5 (excellent).

\subsubsection{ESTOI}
The extended short-time objective intelligibility (ESTOI)~\cite{7539284} is an instrumental metric for evaluating the intelligibility of speech signals, ranging from 0 to 1. The higher ESTOI score indicates a higher intelligibility and better preservation of the speech content.

\subsubsection{CSIG, CBAK, COVL}
We consider three composite mean opinion score (MOS) based measures~\cite{4389058} to further quantify the speech quality, i.e., CSIG (the MOS of signal distortion), CBAK (the intrusiveness of background noise), and COVL (the overall effect). The higher, the better.

\subsubsection{WV-MOS}
The WV-MOS\footnote{https://github.com/AndreevP/wvmos}~\cite{10097255} is a non-intrusive MOS predictor using the fine-tuned wav2vec2.0 model~\cite{NEURIPS2020_92d1e1eb}, which estimates the MOS scores without clean signals.

\subsubsection{ViSQOL}
The virtual speech quality objective listener (ViSQOL)\footnote{https://github.com/google/visqol}~\cite{9123150} utilizes the spectral-temporal similarity between reference and test speech signals to produce a mean opinion score - listening quality objective (MOS-LQO) score.

\subsubsection{LSD}
The log-spectral distance (LSD)~\cite{1162849} is an STFT-domain metric to evaluate the speech SR performance, which calculates the logarithmic distance between the clean and degraded magnitude spectrum as
\begin{align}
{\rm LSD} = \frac{1}{K} \sum_{k=1}^{K}{\sqrt{ \frac{1}{F} \sum_{f=1}^{F}{ {\rm ln}\left( \frac{\tilde{\bm{X}}_m(k,f)^2}{\hat{\tilde{\bm{X}}}_m(k,f)^2} \right)^2
 } }}.
\end{align}
A lower LSD indicates a better SR performance, and 0 means the minimum distance.

\subsubsection{SSIM}
Structural similarity index measure (SSIM)~\cite{1284395} was originally proposed to assess the image quality by comparing local pixel patterns in terms of luminance, contrast, and structure. We compute this measure on the magnitude spectrum to evaluate the speech SR performance.

\subsubsection{WER}
The word error rate (WER) is used to further evaluate the clarity of the enhanced speech signals in combination with a downstream ASR task. We utilize the squeezeformer model~\cite{NEURIPS2022_3ccf6da3} pre-trained by NVIDIA\footnote{\url{https://catalog.ngc.nvidia.com/orgs/nvidia/teams/nemo/models/stt_en_squeezeformer_ctc_xsmall_ls}} for English speech recognition, which is the x-small version with 8.8M parameters.

\subsection{Implementation}
We perform STFT using a Hann window with a length of 512 (32ms) and a shift of 192 (12ms). We chose $\beta_1$ = 0.3 and $\beta_2$ = 0.3 for the amplitude transformation in  \eqref{transform}. 
The numbers of output channels for Conv blocks in the encoder and decoder are \{16, 32, 48, 64\} and \{48, 32, 16, 2 or 1\} (2 for the predictive branch and 1 for the generative branch), respectively. The number of hidden states in the Bi-LSTM is set to 128, and there are 4 heads in the MHSA layer. For the SDE formulation, BBED with $k$ = 2.6, $c$ = 0.51 and $T$ = 0.999 is utilized following~\cite{lay23_interspeech}. 
We adopt an exponential moving average of model weights with a factor of 0.999 for sampling~\cite{song2020improved}.
Our model is trained with the AdamW optimizer
for 200 epochs. The $L_2$ norm for gradient clipping is set to 5.0. The learning rate starts from 1e-3 and decays at a factor of 0.97 every two epochs. Our training is conducted on NVIDIA RTX4080 (16GB memory) and takes one day.

\begin{table*}[]
    \scriptsize
    \centering
    \caption{Speech enhancement results obtained on the WSJ0-UNI dataset in the form of mean $\pm$ standard deviation, where `P' and `G' denote predictive and generative methods, respectively.}
    
    \begin{tabular*}{\textwidth}{@{\extracolsep{\fill}}l|ccc|cccccccc}
        \toprule
         Method & Para. & MACs & Type & PESQ & ESTOI & CSIG & CBAK & COVL & WV-MOS & ViSQOL \\
         \midrule
         \midrule
         Degraded & - & - & - & 2.40~$\pm$~1.26 & 0.81~$\pm$~0.17 & 3.29~$\pm$~1.31 & 2.85~$\pm$~1.00 & 2.88~$\pm$~1.31 & 2.47~$\pm$~1.97 & 2.11~$\pm$~1.13 \\
         \midrule
         Conv-TasNet~\cite{8707065} & 3.4M & 3.2G & P & 2.81~$\pm$~1.13 & 0.87~$\pm$~0.14 & 3.90~$\pm$~0.92 & 3.27~$\pm$~0.88 & 3.45~$\pm$~1.07 & 3.14~$\pm$~1.16 & 2.44~$\pm$~1.08 \\
         MANNER~\cite{9747120} & 24.1M & 8.7G & P & 3.16~$\pm$~1.03 & 0.91~$\pm$~0.10 & 4.42~$\pm$~0.60 & 3.47~$\pm$~0.80 & 3.91~$\pm$~0.88 & 3.49~$\pm$~0.64 & 2.84~$\pm$~1.10 \\
         CMGAN~\cite{10508391} & 1.8M & 31.7G & P & 3.43~$\pm$~0.90 & 0.92~$\pm$~0.09 & 4.46~$\pm$~0.67 & 3.56~$\pm$~0.73 & 4.06~$\pm$~0.82 & 3.69~$\pm$~0.54 & 2.81~$\pm$~1.00 \\
         \midrule
         CDiffuSE~\cite{9746901} & 4.3M & 292.4G & G & 2.20~$\pm$~0.67 & 0.80~$\pm$~0.13 & 3.64~$\pm$~0.75 & 2.74~$\pm$~0.49 & 2.96~$\pm$~0.71 & 3.03~$\pm$~1.18 & 1.78~$\pm$~0.44 \\
         SGMSE+~\cite{10149431} & 65.6M & 8.0T & G & 3.19~$\pm$~1.09 & 0.91~$\pm$~0.10 & 4.18~$\pm$~0.84 & 3.45~$\pm$~0.83 & 3.79~$\pm$~1.02 & 3.76~$\pm$~0.48 & 2.73~$\pm$~1.06 \\
         StoRM~\cite{10180108} & 55.1M & 15.8T & G+P & 3.17~$\pm$~1.09 & 0.91~$\pm$~0.10 & 4.28~$\pm$~0.77 & 3.46~$\pm$~0.86 & 3.84~$\pm$~0.99 & 3.74~$\pm$~0.46 & 2.69~$\pm$~1.05 \\
         UNIVERSE++~\cite{scheibler24_interspeech} & 42.9M & 42.8G & G+P & 3.20~$\pm$~1.01 & 0.91~$\pm$~0.10 & 4.33~$\pm$~0.69 & 3.56~$\pm$~0.78 & 3.88~$\pm$~0.92 & 3.75~$\pm$~0.47 & 2.62~$\pm$~0.99 \\
         \midrule
         PGUSE-P & 2.3M & 5.8G & P & 3.36~$\pm$~0.91 & 0.91~$\pm$~0.09 & 4.43~$\pm$~0.61 & 3.60~$\pm$~0.71 & 4.01~$\pm$~0.82 & 3.63~$\pm$~0.53 & 2.78~$\pm$~1.02 \\
         PGUSE-G & 5.1M & 177.3G & G+P & 3.38~$\pm$~0.95 & \textbf{0.93~$\pm$~0.08} & 4.53~$\pm$~0.59 & 3.63~$\pm$~0.75 & 4.09~$\pm$~0.84 & \textbf{3.80~$\pm$~0.43} & 2.92~$\pm$~0.99 \\
         PGUSE-F & 5.1M & 177.3G & G+P & 3.50~$\pm$~0.89 & \textbf{0.93~$\pm$~0.08} & \textbf{4.59~$\pm$~0.54} & 3.71~$\pm$~0.73 & \textbf{4.18~$\pm$~0.78} & 3.79~$\pm$~0.44 & \textbf{2.95~$\pm$~1.00} \\
         PGUSE-T & 5.1M & 26.3G & G+P & 3.46~$\pm$~0.89 & \textbf{0.93~$\pm$~0.08} & 4.55~$\pm$~0.56 & 3.69~$\pm$~0.72 & 4.14~$\pm$~0.80 & 3.78~$\pm$~0.46 & 2.91~$\pm$~0.95 \\
         PGUSE & 5.1M & 26.3G & G+P & \textbf{3.53~$\pm$~0.87} & \textbf{0.93~$\pm$~0.08} & 4.58~$\pm$~0.53 & \textbf{3.74~$\pm$~0.72} & \textbf{4.18~$\pm$~0.77} & 3.76~$\pm$~0.47 & 2.93~$\pm$~0.98 \\
         \bottomrule
    \end{tabular*}
    \vspace{-1em}
    \label{tab:WSJ0-UNI}
\end{table*}

\subsection{Hyperparameter Search}
\label{sec:search}

We conduct a hyperparameter search on the WSJ0-UNI dataset to find the optimal settings for the output fusion factor $\alpha$ in  \eqref{fusion}, the number of reverse steps $N$, and the reverse start time $T_{rs}$ of the truncated diffusion. To visualize the results, we employ both intrusive PESQ and non-intrusive WV-MOS metrics, where the former assesses the fidelity of the reconstruction relative to a reference signal, while the latter allows the speech quality assessment of a realization on the manifold of clean speech.

\subsubsection{Output fusion factor $\alpha$}
The factor $\alpha$ regulates the ratio of predictive and generative components in the output, and the performance curves in terms of $\alpha$ are presented in Fig. \ref{fig:alpha}, with $N$ being fixed to 25. It can be seen the generative method ($\alpha$ = 0) outperforms the predictive method ($\alpha$ = 1) in terms of WV-MOS, as the former focuses on generating plausible samples, while the latter will encounter deviations when predicting deterministic reference. The PESQ score reaches its highest when $\alpha$ = 0.5, indicating that output fusion can enhance the reconstruction accuracy of the reference. We thus choose $\alpha$ = 0.4 as a trade-off between the two metrics in the sequel.

\subsubsection{Number of reverse steps $N$}
Fig. \ref{fig:N} shows the performance in terms of $N$ by fixing $\alpha$ to 0.4. It is clear that 25 steps are enough and both metrics show no further increase with $N>25$. Therefore, we set $N=25$ in the sequel, which leads to a step width of $\Delta t = 1/N= 0.04$.

\subsubsection{Reverse start time $T_{rs}$}
In Fig. \ref{fig:Trs}, we utilize the truncated diffusion scheme and depict the impact of $T_{rs}$ on the performance with $\alpha$ = 0.4 and $\Delta t$ = 0.04. 
As a larger $T_{rs}$ causes more reverse steps and a greater complexity, we choose $T_{rs}$ = 0.12 with 3 reverse steps. Notably, it achieves remarkable metric scores of approximately 3.51 in PESQ and 3.76 in WV-MOS with even one reverse step, indicating the effectiveness of the truncated diffusion approach.

\subsection{Comparison Models}

Our proposed PGUSE is objectively compared with other SOTA SE methods, including predictive approaches (Conv-TasNet~\cite{8707065}, MANNER~\cite{9747120}, CMGAN~\cite{10508391}) and generative approaches (CDiffuSE~\cite{9746901}, SGMSE+~\cite{10149431}, StoRM~\cite{10180108}, UNIVERSE++~\cite{scheibler24_interspeech}) described in detail below. We re-trained all models in experiments, unless stated elsewhere.
\subsubsection{Conv-TasNet}
An end-to-end neural network designed for speech separation, which leverages temporal convolutional networks to effectively separate mixed audio signals by masking the learned representation of the mixture.
\subsubsection{MANNER}
A time-domain SE model using U-net based encoder-decoder architecture. It employs multi-view attention to capture full information from the signal, efficiently addressing both channel and long-sequential features.
\subsubsection{CMGAN}
A time-frequency domain model using dual-path conformer blocks~\cite{gulati20_interspeech} to encode both magnitude and complex spectrum information. A metric discriminator~\cite{pmlr-v97-fu19b} is trained to alleviate the mismatch between the speech quality and optimization objectives.
\subsubsection{CDiffuSE}
This method generalizes discrete-time diffusion by incorporating the observed noisy data into the model, resulting in a conditional diffusion process in the time domain.
\subsubsection{SGMSE+}
A score-based diffusion model defined in the complex spectrum domain. It adopts NCSN++ network~\cite{song2021scorebased} and follows the OUVE SDE formulation.
\subsubsection{StoRM}
The stochastic regeneration model (StoRM) utilizes a predictive model for initial recovery and a generative model for refinement, on the basis of the SGMSE+ framework.
\subsubsection{UNIVERSE++}
A score-based time-domain diffusion model designed for USE, where the condition network encodes the degraded speech and the score network estimates score functions. Multiple losses constrain the condition network to predict the clean speech, but the predictive results lack further integration with generative outcomes.

\begin{figure}
    \centering
    \includegraphics[width=0.85\columnwidth]{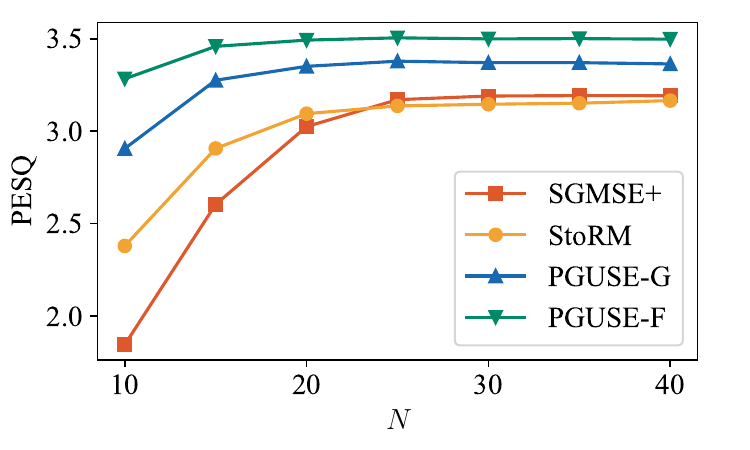}
    \caption{The PESQ of SGMSE+, StoRM, PGUSE-G and PGUSE-F in terms of different reverse steps.}
    \vspace{-1em}
    \label{fig:comparison_N}
\end{figure}

\section{Results and Discussions}

\begin{table*}[]
    \centering
    \caption{Speech denoising results obtained on the VBDMD test set under different training conditions in the form of mean $\pm$ standard deviation. Models marked with $\dagger$ are pre-trained by authors using the same training data. }

    \begin{tabular*}{\textwidth}{@{\extracolsep{\fill}}l|cc|cccccccc}
        \toprule
         Method & Type & Training set & PESQ & ESTOI & CSIG & CBAK & COVL & WV-MOS & ViSQOL \\
         \midrule
         \midrule
         Degraded & - & - & 1.98~$\pm$~0.76 & 0.79~$\pm$~0.15 & 3.48~$\pm$~0.83 & 2.54~$\pm$~0.64 & 2.74~$\pm$~0.80 & 3.00~$\pm$~1.25 & 2.09~$\pm$~0.92 \\
         \midrule
         Conv-TasNet~\cite{8707065} & P & VBDMD & 2.56~$\pm$~0.64 & 0.85~$\pm$~0.10 & 3.89~$\pm$~0.68 & 3.45~$\pm$~0.49 & 3.27~$\pm$~0.66 & 4.21~$\pm$~0.40 & 2.55~$\pm$~0.83 \\
         MANNER\textsuperscript{$\dagger$}~\cite{9747120} & P & VBDMD & 3.20~$\pm$~0.62 & 0.87~$\pm$~0.09 & 4.54~$\pm$~0.50 & 3.72~$\pm$~0.47 & 3.94~$\pm$~0.60 & 4.36~$\pm$~0.30 & 2.93~$\pm$~0.88 \\
         CMGAN~\cite{10508391} & P & VBDMD & \textbf{3.38~$\pm$~0.63} & \textbf{0.89~$\pm$~0.09} & 4.60~$\pm$~0.50 & \textbf{3.87~$\pm$~0.49} & \textbf{4.08~$\pm$~0.62} & 4.36~$\pm$~0.31 & \textbf{3.16~$\pm$~0.82} \\
         PGUSE-P & P & VBDMD & 3.20~$\pm$~0.69 & 0.88~$\pm$~0.09 & 4.48~$\pm$~0.56 & 3.73~$\pm$~0.48 & 3.91~$\pm$~0.66 & 4.33~$\pm$~0.35 & 3.00~$\pm$~0.88 \\
         \midrule
         CDiffuSE\textsuperscript{$\dagger$}~\cite{9746901} & G & VBDMD & 2.48~$\pm$~0.55 & 0.79~$\pm$~0.11 & 3.77~$\pm$~0.55 & 3.03~$\pm$~0.43 & 3.15~$\pm$~0.55 & 3.62~$\pm$~0.72 & 2.03~$\pm$~0.56 \\
         SGMSE+\textsuperscript{$\dagger$}~\cite{10149431} & G & VBDMD & 2.88~$\pm$~0.61 & 0.86~$\pm$~0.10 & 4.24~$\pm$~0.62 & 3.48~$\pm$~0.46 & 3.60~$\pm$~0.62 & 4.23~$\pm$~0.33 & 2.78~$\pm$~0.85 \\
         StoRM\textsuperscript{$\dagger$}~\cite{10180108} & G+P & VBDMD & 2.85~$\pm$~0.63 & 0.87~$\pm$~0.10 & 4.18~$\pm$~0.62 & 3.53~$\pm$~0.48 & 3.56~$\pm$~0.63 & 4.28~$\pm$~0.35 & 2.84~$\pm$~0.87 \\
         UNIVERSE++~\cite{scheibler24_interspeech} & G+P & VBDMD & 3.03~$\pm$~0.64 & 0.87~$\pm$~0.10 & 4.38~$\pm$~0.57 & 3.62~$\pm$~0.48 & 3.76~$\pm$~0.62 & \textbf{4.41~$\pm$~0.30} & 2.87~$\pm$~0.86 \\
         PGUSE & G+P & VBDMD & 3.30~$\pm$~0.67 & 0.88~$\pm$~0.09 & \textbf{4.63~$\pm$~0.50} & 3.79~$\pm$~0.48 & 4.05~$\pm$~0.63 & 4.34~$\pm$~0.32 & 3.10~$\pm$~0.87 \\
         \midrule
         \midrule
         Conv-TasNet~\cite{8707065} & P & WSJ0-UNI & 2.20~$\pm$~0.63 & 0.79~$\pm$~0.13 & 3.02~$\pm$~0.91 & 2.75~$\pm$~0.39 & 2.64~$\pm$~0.77 & 3.41~$\pm$~0.78 & 1.71~$\pm$~0.66 \\
         MANNER~\cite{9747120} & P & WSJ0-UNI & 2.51~$\pm$~0.60 & 0.83~$\pm$~0.11 & 3.24~$\pm$~0.84 & 2.90~$\pm$~0.35 & 2.91~$\pm$~0.71 & 3.76~$\pm$~0.59 & 1.95~$\pm$~0.69 \\
         CMGAN~\cite{10508391} & P & WSJ0-UNI & \textbf{2.69~$\pm$~0.59} & 0.85~$\pm$~0.10 & 3.44~$\pm$~0.81 & 2.96~$\pm$~0.35 & 3.10~$\pm$~0.69 & 4.13~$\pm$~0.41 & 1.91~$\pm$~0.77 \\
         {PGUSE-P} & P & WSJ0-UNI & 2.60~$\pm$~0.54 & 0.85~$\pm$~0.10 & 3.51~$\pm$~0.65 & 2.97~$\pm$~0.32 & 3.09~$\pm$~0.59 & 4.15~$\pm$~0.38 & 2.13~$\pm$~0.78 \\
         \midrule
         CDiffuSE~\cite{9746901} & G & WSJ0-UNI & 1.71~$\pm$~0.39 & 0.75~$\pm$~0.13 & 2.73~$\pm$~0.44 & 2.29~$\pm$~0.38 & 2.22~$\pm$~0.42 & 2.64~$\pm$~1.05 & 1.54~$\pm$~0.51 \\
         SGMSE+~\cite{10149431} & G & WSJ0-UNI & 2.41~$\pm$~0.66 & 0.84~$\pm$~0.11 & 3.50~$\pm$~0.74 & 2.87~$\pm$~0.44 & 2.98~$\pm$~0.70 & 3.92~$\pm$~0.51 & \textbf{2.19~$\pm$~0.78} \\
         StoRM~\cite{10180108} & G+P & WSJ0-UNI & 2.37~$\pm$~0.58 & 0.82~$\pm$~0.11 & 2.90~$\pm$~0.83 & 2.76~$\pm$~0.36 & 2.67~$\pm$~0.68 & 3.99~$\pm$~0.40 & 1.69~$\pm$~0.66 \\
         UNIVERSE++~\cite{scheibler24_interspeech} & G+P & WSJ0-UNI & 2.46~$\pm$~0.61 & 0.82~$\pm$~0.11 & 3.50~$\pm$~0.63 & 2.85~$\pm$~0.37 & 3.01~$\pm$~0.62 & 4.06~$\pm$~0.41 & 1.94~$\pm$~0.69 \\
         PGUSE & G+P & WSJ0-UNI & \textbf{2.69~$\pm$~0.56} & \textbf{0.86~$\pm$~0.10} & \textbf{3.73~$\pm$~0.60} & \textbf{3.01~$\pm$~0.34} & \textbf{3.24~$\pm$~0.58} & \textbf{4.20~$\pm$~0.39} & \textbf{2.19~$\pm$~0.80} \\
         \bottomrule
    \end{tabular*}
    \vspace{-1em}
    \label{tab:VBDMD}
\end{table*}

\subsection{Universal Speech Enhancement (USE)}
First, in Table \ref{tab:WSJ0-UNI} we report the USE performance obtained on the WSJ0-UNI dataset. Our PGUSE is compared with its variants and selected baselines in terms of the parameter amount (Para.), multiply-accumulate operations (MACs)\footnote{https://github.com/sovrasov/flops-counter.pytorch} and the aforementioned speech quality measures. For the diffusion-based generative models, we report the MACs in the whole inference process, which usually involves several steps as described in the corresponding literature.

As the proposed PGUSE framework is a compound of typical generative and predictive models, we can selectively pick results from each branch or their fusion. 
To show this, we compare PGUSE with its variants at the bottom of Table \ref{tab:WSJ0-UNI}, where PGUSE-P indicates the results of the predictive branch, PGUSE-G the generative branch, PGUSE-F only the output fusion, PGUSE-T only the truncated diffusion scheme, and PGUSE both the two schemes, respectively.
All variants utilize the phase estimated from the predictive branch.
It can be seen that PGUSE-P has a smaller model size and fewer MACs than PGUSE-G, since the former involves only one call of the predictive branch, while the latter requires 25 reverse steps as claimed in Section \ref{sec:search}. PGUSE-G surpasses PGUSE-P in terms of all metrics, especially for WV-MOS and ViSQOL. 
This verifies the potential of generative approaches for the USE task, which suffers from the damaged speech information in the degraded signals.
PGUSE-F can improve most metrics by fusing predictive and generative results, showing a certain degree of complementarity. PGUSE-T shows improvements over PGUSE-G with less computational complexity, 
because predictive results can narrow the gap between diffusion states and target distributions while reducing reverse steps.
PGUSE further improves the performance by combining the output fusion and truncated diffusion scheme, although with a slight decrease in WV-MOS and ViSQOL.

The comparison with predictive baselines demonstrates that the proposed PGUSE achieves improvements for all metrics. Compared to Conv-TasNet which has a minimum computational complexity, our model shows an obvious superiority in performance. In the comparison with the SOTA predictive method CMGAN, our PGUSE still works better, although CMGAN adopts a PESQ discriminator for special optimization. Compared to the diffusion-based generative methods, PGUSE has a better performance with a lighter computational burden, indicating the efficiency and effectiveness of the truncated diffusion scheme. 
We can see that the generative approaches generally outperform the predictive methods in terms of non-intrusive WV-MOS, but perform relatively poorer for other intrusive metrics. This is because predictive methods optimize certain point-wise loss functions between the estimated speech and a clean reference, while the generative methods learn to model the inherent characteristics of speech signals. 
Fig.~\ref{fig:comparison_N} visualizes the PESQ in terms of different reverse diffusion steps, where our PGUSE-G and PGUSE-F obviously outperform SGMSE+ and StoRM. We observe that introducing predictive modeling helps maintain performance with fewer reverse steps, as seen in the comparison from StoRM to SGMSE+ and from PGUSE-F to PGUSE-G.
In summary, our PGUSE integrates the advantages of predictive and generative learning to achieve precise reconstruction and good speech naturalness, all while maintaining a lighter computational overhead, thus establishing a new benchmark of diffusion-based models.

\subsection{Speech Denoising on VBDMD}
Second, results obtained on the VBDMD test set are presented in Table~\ref{tab:VBDMD}. This dataset only involves additive noise distortion to verify the denoising ability of models. Observing the match condition in the upper half of Table \ref{tab:VBDMD}, where training samples are from VBDMD, we find that CMGAN outperforms other models in terms of most metrics. This is due to the fact that predictive methods are competent for the conventional denoising task, as there are enough speech clues for regression learning, unless under extremely low SNR conditions. 
Our PGUSE considers both predictive and generative modeling, surpassing other generative baselines and narrowing the gap between generative approaches and advanced predictive models on the VBDMD benchmark. We also report the results of PGUSE-P, which is inferior to that of PGUSE. This indicates that generative modeling can improve the upper limit of predictive methods, even in the context of straightforward denoising task.

The bottom half of Table \ref{tab:VBDMD} compares the results under a mismatch condition, where models are trained on the WSJ0-UNI dataset. 
This cross-dataset evaluation is to show the transferability of the USE models to the denoising task with unseen data distribution.
We observe that the proposed PGUSE shows superiority in terms of all metrics, revealing an excellent generalization ability. We also observe that StoRM exhibits a performance degradation when compared to SGMSE+, since the mismatch condition brings instability to the cascade framework of predictive model and generative refinement. In contrast, PGUSE adopts a parallel structure and shows a more stable performance in this more challenging case.

\begin{table}[]
    \centering
    \caption{Speech dereverberation performance obtained on VBDMD-REVERB with models trained on WSJ0-UNI.}

    \resizebox{\columnwidth}{!}{
        \begin{tabular}{l|ccccc}
        \toprule
        Method & PESQ & ESTOI & COVL & WV-MOS & ViSQOL \\
        \midrule
        \midrule
        Degraded & 1.71 & 0.85 & 2.68 & 3.86 & 2.87 \\
        \midrule
        Conv-TasNet~\cite{8707065} & 2.27 & 0.86 & 2.94 & 3.69 & 2.59 \\
        MANNER~\cite{9747120} & 2.67 & 0.91 & 3.29 & 3.92 & 2.94 \\
        CMGAN~\cite{10508391} & 3.14 & 0.93 & 3.60 & 4.39 & 3.31 \\
        \midrule
        CDiffuSE~\cite{9746901} & 2.04 & 0.81 & 2.71 & 3.75 & 2.32 \\
        SGMSE+~\cite{10149431} & 3.24 & 0.93 & \textbf{3.88} & 4.41 & \textbf{3.84} \\
        StoRM~\cite{10180108} & 3.12 & 0.93 & 3.67 & 4.34 & 2.80 \\
        UNIVERSE++~\cite{scheibler24_interspeech} & 2.96 & 0.90 & 3.58 & 4.27 & 3.20 \\
        \midrule
        PGUSE & \textbf{3.37} & \textbf{0.94} & \textbf{3.88} & \textbf{4.46} & 3.64 \\
        \bottomrule
        \end{tabular}
    }
    \label{tab:VBDMD-REVERB}
\end{table}

\subsection{Speech Dereverberation on VBDMD-REVERB}
Third, we evaluate the speech dereverberation performance of our PGUSE model in comparison with other baselines on the VBDMD-REVERB dataset in Table~\ref{tab:VBDMD-REVERB}. It can be seen that SGMSE+ performs better than CMGAN, indicating the effectiveness of diffusion models in detecting the correlation of particular time-frequency bins with corresponding dry speech areas. Since the reverberant signal originates from the dry source, which causes less speech uncertainty than other distortions, the vocalized artifacts observed in the denoising task can be reduced~\cite{10149431}. More importantly, the proposed PGUSE model still exhibits the leading performance in terms of most metrics, showing a strong applicability to dereverberation and robustness against unseen data.


\begin{table}[]
    \centering
    \caption{Speech SR (8 kHz $\rightarrow$ 16 kHz) results obtained on VBDMD-SR with models trained on WSJ0-UNI.}

    \resizebox{\columnwidth}{!}{
        \begin{tabular}{l|ccccc}
        \toprule
        Method & LSD$\downarrow$ & SSIM$\uparrow$ & PESQ$\uparrow$ & CSIG$\uparrow$ & COVL$\uparrow$ \\
        \midrule
        \midrule
        Degraded & 5.13 & 0.77 & \textbf{4.26} & 1.68 & 3.03 \\
        \midrule
        Conv-TasNet~\cite{8707065} & 3.18 & 0.78 & 3.48 & 3.30 & 3.45 \\
        MANNER~\cite{9747120} & 2.90 & 0.81 & 3.03 & 3.90 & 3.52 \\
        CMGAN~\cite{10508391} & 2.62 & 0.82 & 3.73 & 3.54 & 3.69 \\
        \midrule
        CDiffuSE~\cite{9746901} & 3.15 & 0.71 & 2.66 & 3.42 & 3.08 \\
        SGMSE+~\cite{10149431} & 2.69 & 0.85 & 3.84 & 3.86 & 3.91 \\
        StoRM~\cite{10180108} & 2.88 & 0.76 & 2.89 & 3.50 & 3.24 \\
        UNIVERSE++~\cite{scheibler24_interspeech} & 2.67 & 0.77 & 3.01 & 3.93 & 3.52 \\
        \midrule
        PGUSE & 2.16 & 0.87 & 3.81 & 4.10 & 4.00 \\
        PGUSE-LFR & \textbf{2.03} & \textbf{0.88} & 4.09 & \textbf{4.41} & \textbf{4.32} \\
        \bottomrule
        \end{tabular}
    }
    \vspace{-1em}
    \label{tab:VBDMD-SR}
\end{table}

\subsection{Speech Super-Resolution (SR)}

Furthermore, we compare the speech SR performance on the VBDMD-SR dataset using the models trained on the WSJ0-UNI dataset in Table \ref{tab:VBDMD-SR}. Compared with other methods, our PGUSE maintains a leading position. The LSD and SSIM metrics indicate that PGUSE effectively restores the spectral structure more accurately, while the CSIG and COVL metrics demonstrate improvements in the perceptual speech quality. It is interesting that the degraded speech with a limited bandwidth achieves the highest PESQ score, and all models suffer from a decline. This can happen because the PESQ is not specifically designed for speech SR evaluation and low frequencies dominate the PESQ measure, which significantly impacts human hearing. The processes of models will inevitably introduce errors to the low-frequency regions, leading to a degradation in PESQ. 
For this, speech SR methods usually perform a lower-frequencies replacement (LFR) operation to improve the performance~\cite{liu22x_interspeech} by reusing the low-frequency components of band-limited signals. We observe that the PGUSE with LFR can clearly produce better results, though the PESQ is still inferior to that of the degraded speech.

\begin{figure}
    \centering
    \vspace{-0.2cm}
    \includegraphics[width=0.95\columnwidth]{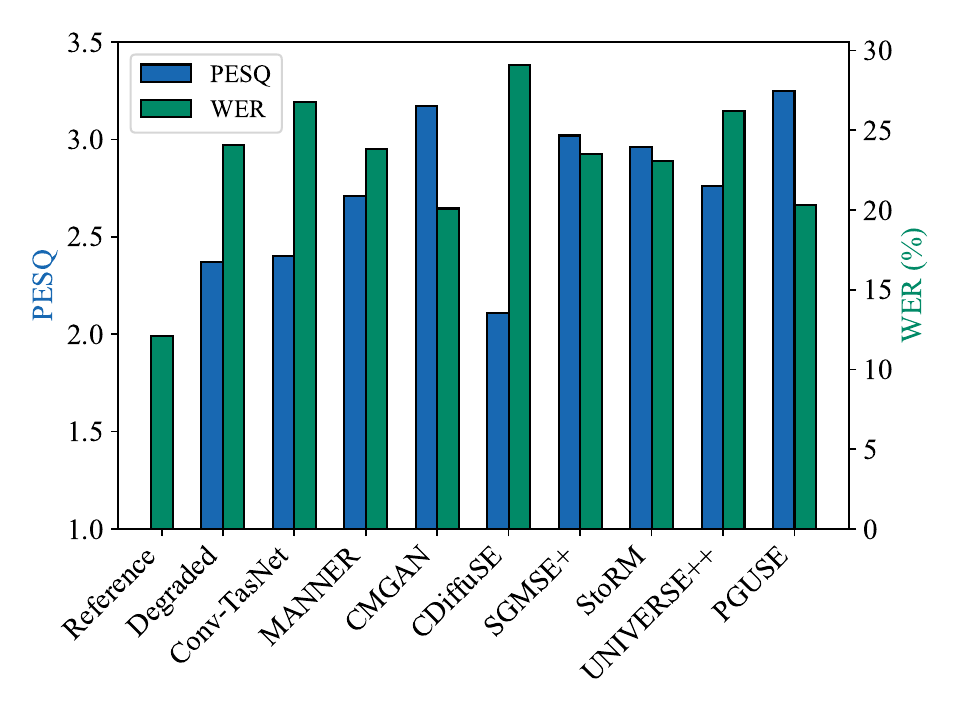}
    \vspace{-0.7em}
    \caption{SE and ASR results on the TIMIT-UNI dataset.}
    \label{fig:wer}
\end{figure}

\subsection{Application to downstream ASR}
In addition, we simultaneously show the SE and ASR results on the TIMIT-UNI dataset in Fig.~\ref{fig:wer}. Our PGUSE achieves the highest PESQ score and reduces the overall WER over the degraded utterances. The generative baselines demonstrate a higher WER  compared to that of the predictive CMGAN, which can be attributed to the vocalizing artifacts and phonetic confusions arising from generative behaviors under highly adverse conditions~\cite{10180108}. StoRM achieves a better ASR performance than SGMSE+, indicating the stochastic regeneration approach can correct some artifacts. The proposed PGUSE can efficiently combine predictive and generative modeling capacities to improve the reconstruction accuracy and reduce artifacts, achieving a WER comparable to the advanced predictive CMGAN model.

\subsection{Ablation Study}

\begin{table}[]
    \centering
    \caption{Ablation study on the WSJ0-UNI dataset, including ablation of network components and formalism.}

    \resizebox{\columnwidth}{!}{
        \begin{tabular}{l|ccccc}
        \toprule
        Method & PESQ & CSIG & CBAK & COVL & WV-MOS \\
        \midrule
        \midrule
        PGUSE & \textbf{3.53} & \textbf{4.58} & \textbf{3.74} & \textbf{4.18} & \textbf{3.76} \\
        \midrule
        w/o Interaction & 3.51 & 4.57 & 3.72 & 4.17 & \textbf{3.76} \\
        w/o Sub-band Conv & 3.50 & \textbf{4.58} & 3.71 & 4.17 & 3.71 \\
        w/o MHSA & 3.47 & 4.52 & 3.59 & 4.12 & 3.71 \\
        w/o ConvGLU & 3.49 & 4.56 & 3.69 & 4.15 & 3.72\\
        \midrule
        OUVE & 3.46 & 4.50 & 3.69 & 4.10 & 3.65 \\
        Complex & 3.27 & 4.40 & 3.49 & 3.95 & 3.55 \\
        Degraded Phase & 3.47 & 4.54 & 3.56 & 4.13 & 3.73 \\
        \bottomrule
        \end{tabular}
    }
    \vspace{-1em}
    \label{tab:ablation}
\end{table}

Finally, we carry out ablation studies on the WSJ0-UNI to analyze the impact of different components in the proposed PGUSE model in Table \ref{tab:ablation}. For the network structure,  we first replace the interaction module with a simple addition of features from the predictive and generative branches (i.e., w/o Interaction). This causes a slight performance drop, indicating the effectiveness of this module in transferring valuable information. Substituting the sub-band Conv with a normal convolution layer with a stride of 2 also leads to a performance drop, confirming the importance of extracting band-aware features. The removal of the MHSA layer or ConvGLU module further validates that the attention mechanism enhances the long-term modeling capacity and the ConvGLU is beneficial for aggregating inter-channel information.

From the perspective of formalism, we observe that the OUVE formulation (with $k$ = 10, $c$ = 0.01, $\gamma$ = 1.5 as in~\cite{lay23_interspeech}) results in a degraded SE performance. 
Several factors may contribute to BBED's superior performance over OUVE, such as reduced prior mismatch in the reverse process or higher variance in the SDE evolution, which could potentially help to generate better speech estimates~\cite{lay23_interspeech}. The detailed analysis of OUVE and BBED SDE formulation is beyond the scope of this work. When performing diffusion on the real and imaginary parts of the complex spectrum without modifying the predictive branch (denoted as ``Complex"), we note a clear degradation in performance. This confirms the advantage of operating diffusion in the magnitude STFT domain, which displays clearer patterns. In contrast, the complex spectrum contains numerous unstructured textures in the image sense that might hinder the denoising process during score function estimation. Furthermore, combining the enhanced magnitude with the noisy phase decreases the performance, underscoring the usefulness of the predictive branch for phase enhancement. The complex spectral mapping can thus compensate for the missing phase estimation in the magnitude diffusion process.

\section{Conclusion}
\label{sec:conc}

In this work, we proposed a joint predictive and generative modeling approach for USE (PGUSE). The proposed PGUSE model comprises two parallel branches, where the predictive branch performs complex spectral mapping to directly predict the clean complex spectrum, and the generative branch estimates score functions within a score-based diffusion process to generate candidates in the magnitude STFT domain. Our well-designed neural network ensures robust modeling capabilities for time and frequency patterns, supporting joint predictive and generative training. 
We employed an output fusion scheme to effectively complement the predictive and generative results and adapted the truncated diffusion technique to reduce the number of reverse steps.
We evaluated the proposed PGUSE model across several tasks (e.g., USE, speech denoising, dereveberation, ASR) to show its robustness and capacity. Compared to predictive baselines, our model achieves superior performance for the USE task; compared to generative baselines, our approach delivers a higher reconstruction quality with a significantly lighter computational burden, promoting the practical applications of the diffusion-based models for SE. The combination of predictive and generative methods therefore shows a stronger potential. Future work could consider more distortion types to better simulate real-life situations, and efforts can be made to further optimize the model for real-time processing on low-resource edge devices. The proposed method also occasionally suffers from artifacts similarly to existing models, which might stem from inadequate accounting for phase information in the magnitude estimation process, particularly in high-frequency bands where small non-zero magnitude estimates might be paired with imprecise phase estimates. This requires the incorporation of phase information into the diffusion path while ensuring the ease of estimating Gaussian noise components from diffusion states in the future.

\section{Acknowledgment}
Thanks to the associate editor and anonymous reviewers for their insightful comments that help to improve the quality of this paper. The reproducible code and audio examples are available at https://hyyan2k.github.io/PGUSE. 

\bibliographystyle{ieeetr}
\bibliography{refs}

\begin{IEEEbiography}[{\includegraphics[width=1in,height=1.25in,clip,keepaspectratio]{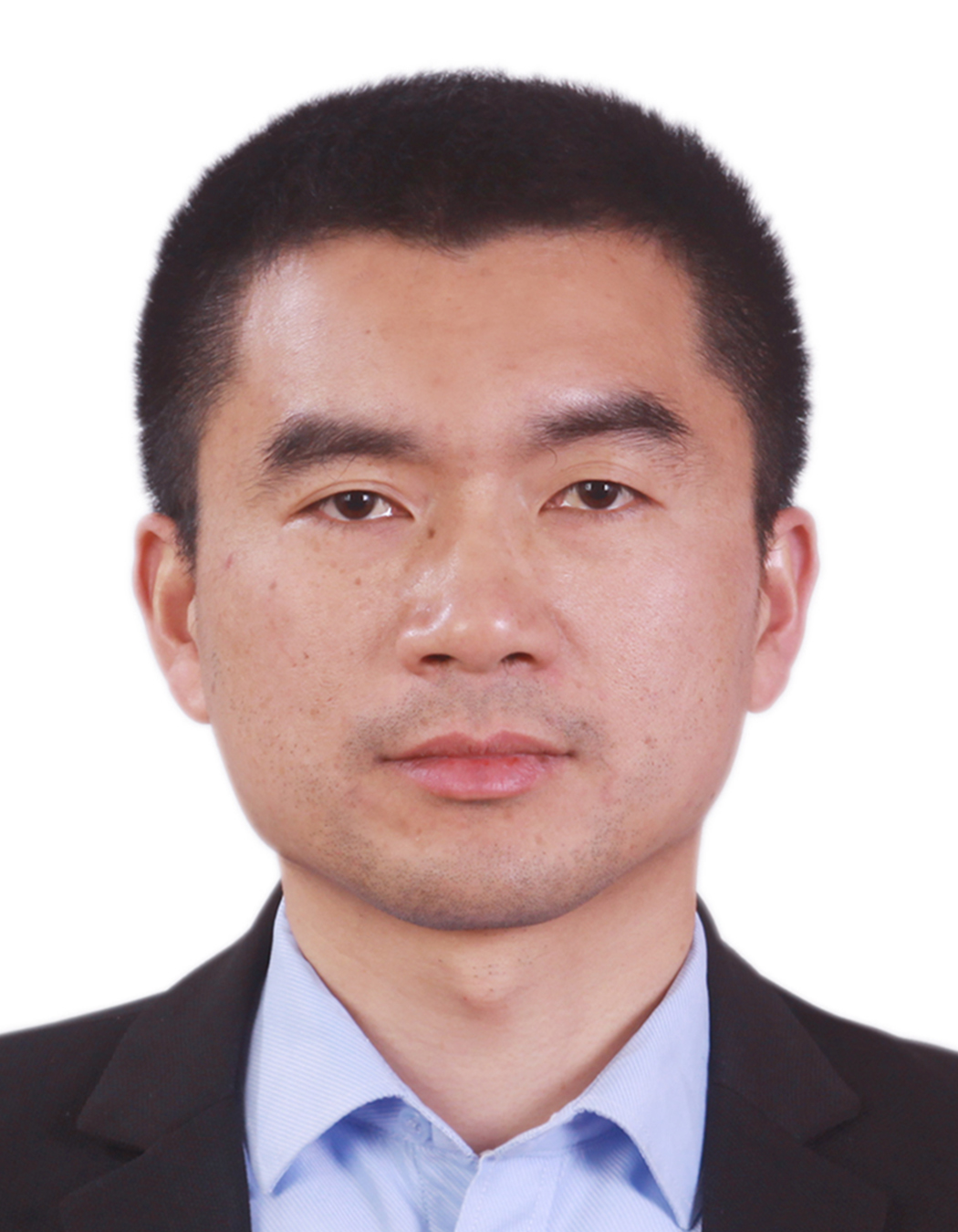}}]{Jie Zhang} (Senior Member, IEEE) received the B.Sc. (with honors from Yunnan University, Yunnan, China),  M.Sc. (with honors from Peking University, Beijing, China), and Ph.D. degrees (from Delft University of Technology (TU Delft), Delft, The Netherlands) in electrical engineering, in 2012, 2015 and 2020, respectively. He is currently an Associate Professor in the National Engineering Research Center for Speech and Language Information Processing (NERC-SLIP), Faculty of Information Science and Technology, University of Science and Technology of China (USTC), Hefei, China. His team won several academic champions, e.g., ChineseAAD of ISCSLP2024, IJCAI-DADA2023 (Deepfake Audio Detection and Analysis), IWSLT2023 (offline and dialect tracks), the second DiCOVA of ICASSP2022 (diagnosing COVID-19 using acoustics) competition, NIST-OpenASR2021, L3DAS23 of ICASSP2023. He received the Best Student Paper Award for his publication at the 10th IEEE Sensor Array and Multichannel Signal Processing Workshop (SAM 2018) in Sheffield, UK. His current research interests include single/multi-microphone speech processing, binaural hearing aids, brain-assisted speech perception and wireless (acoustic) sensor networks. He serves as an Associate Editor for IEEE Transactions on Audio, Speech and Language Processing (TASLPRO).
\end{IEEEbiography}

\begin{IEEEbiography}
[{\includegraphics[width=1in,height=1.25in,clip,keepaspectratio]{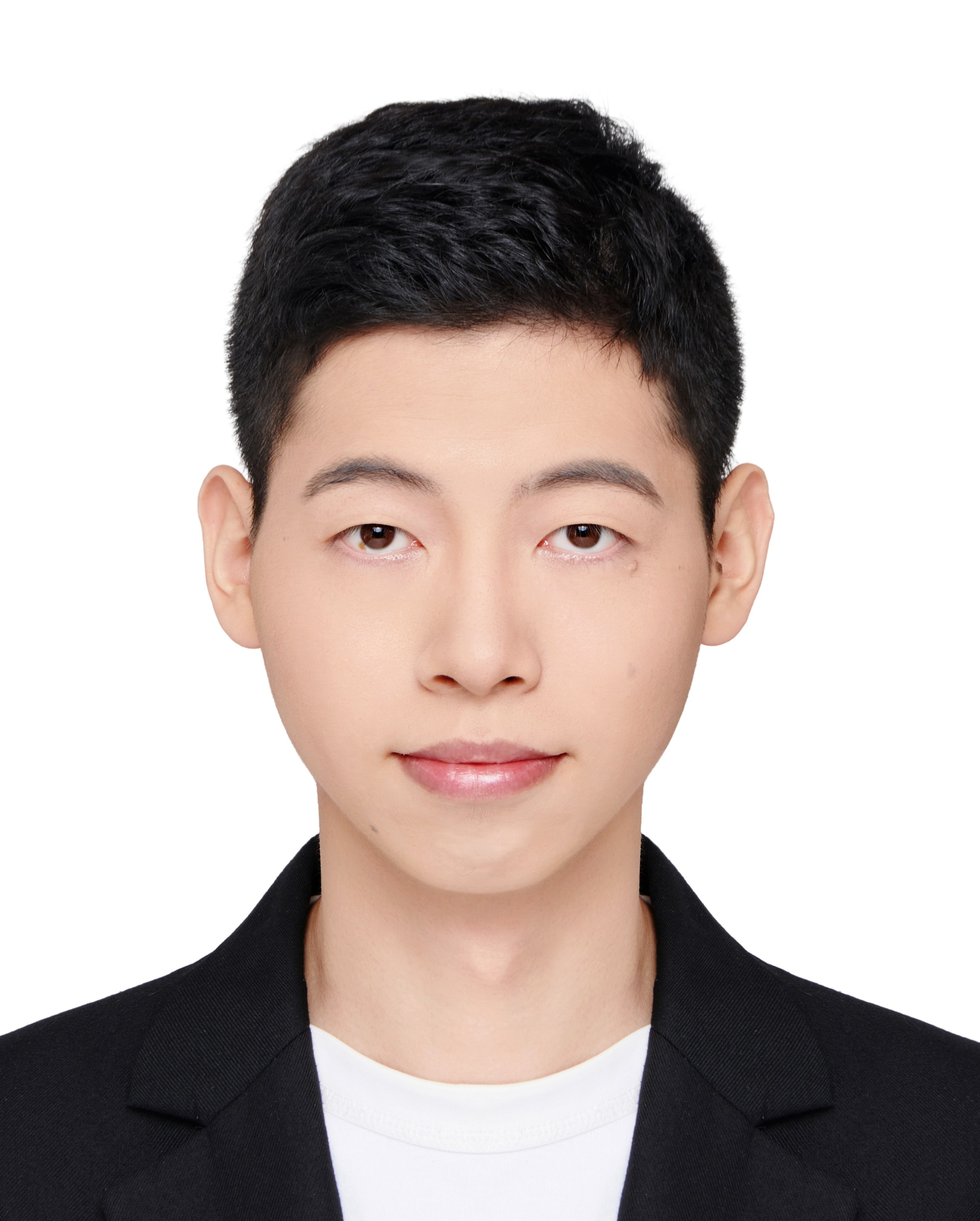}}]{Haoyin Yan} received the B.Sc. degree in electrical information engineering from University of Science and Technology of China (USTC), Hefei, China, in June 2023. He is currently working toward the M.Sc. degree in information and communication engineering with the National Engineering Research Center for Speech and Language Information Processing (NERC-SLIP), Faculty of Information Science and Technology, USTC, Hefei, China. His research interests include speech and audio signal processing, including single/multi-microphone speech enhancement, target speaker extraction, and speech generation.
\end{IEEEbiography}

\begin{IEEEbiography}[{\includegraphics[width=1in,height=1.25in,clip,keepaspectratio]{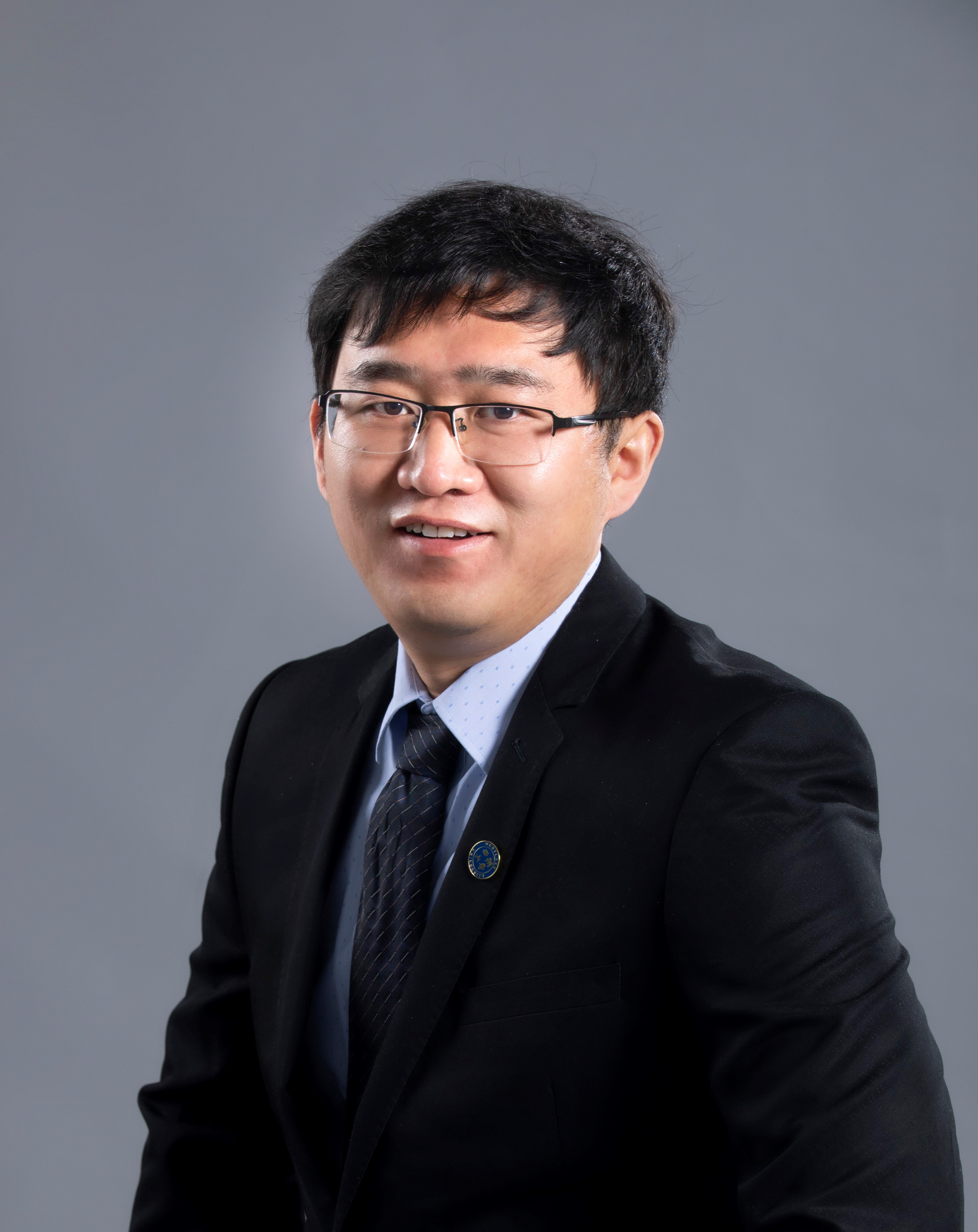}}]{Xiaofei Li} received the Ph.D. degree from Peking University, Beijing, China, in July 2013. From 2014 to 2016, he was with INRIA Grenoble Rhône-Alpes, France, as a Postdoctoral Researcher, and as a Starting Research Scientist from 2016 to 2019. He is currently an Assistant Professor with Westlake University, Hangzhou, China. His research interests include the field of acoustic, audio and speech signal processing, including the topics of speech denoising, dereverberation, separation and localization, sound/speech semisupervised learning and unsupervised pre-training, and sound field reproduction and personal sound zone.
\end{IEEEbiography}
\end{document}